\documentclass[sigconf]{acmart}

\usepackage[utf8]{inputenc}
\usepackage[T1]{fontenc}
\usepackage{booktabs}
\usepackage{graphicx}
\usepackage{multirow}
\usepackage{subcaption}
\usepackage{enumitem}
\usepackage{amsmath}
\usepackage{amsfonts}
\usepackage{xcolor}
\usepackage{pgfplots}
\pgfplotsset{compat=1.18}
\usepackage[most]{tcolorbox}
\usepackage{microtype}
\usepackage{url}
\usepackage{pifont}
\usepackage[capitalise,noabbrev,nameinlink]{cleveref}
\usepackage{listings}
\usepackage{xcolor}

\lstdefinelanguage{json}{
    basicstyle=\footnotesize\ttfamily,
    keywordstyle=\color{blue!70}\bfseries,
    stringstyle=\color{teal},
    commentstyle=\color{gray},
    morestring=[b]",
    morecomment=[l]{//},
    literate=
      *{0}{{{\color{orange!80}0}}}{1}
       {1}{{{\color{orange!80}1}}}{1}
       {...}{{{\color{gray}...}}}{3},
    frame=single,
    framerule=0.4pt,
    rulecolor=\color{gray!40},
    backgroundcolor=\color{gray!6},
    breaklines=true,
    showstringspaces=false,
}

\renewcommand{\arraystretch}{1.15}

\copyrightyear{2026}
\acmYear{2026}
\setcopyright{cc}
\setcctype{by-nc-nd}
\acmConference[CCS '26]{Proceedings of the 2026 ACM SIGSAC Conference on Computer and Communications Security}{November 15--19, 2026}{The Hague, Netherlands}
\acmBooktitle{Proceedings of the 2026 ACM SIGSAC Conference on Computer and Communications Security (CCS '26), November 15--19, 2026, The Hague, Netherlands}
\acmDOI{10.1145/3830454.3846719}
\acmISBN{979-8-4007-2871-6/2026/11}
\renewcommand{\shortauthors}{Yixuan Liu et al.}

\ccsdesc[500]{Security and privacy}
\ccsdesc[500]{Computing methodologies~Artificial intelligence}
\ccsdesc[300]{Security and privacy~Systems security}

\keywords{Linux privilege escalation, LLM agents, security benchmark, threat measurement}

\title{PrivEscalate: Measuring and Augmenting the Threat of LLM-Automated Linux Privilege Escalation}

\author{Yixuan Liu}
\affiliation{%
  \institution{Nanyang Technological University}
  \city{Singapore}
  \country{Singapore}
}
\email{LIUY0255@e.ntu.edu.sg}

\author{Zilong Zhen}
\affiliation{%
  \institution{Nanyang Technological University}
  \city{Singapore}
  \country{Singapore}
}
\email{ZILONG001@e.ntu.edu.sg}

\author{Yin Wu}
\affiliation{%
  \institution{Xi'an Jiaotong University}
  \city{Xi'an}
  \country{China}
}
\email{wuyin@stu.xjtu.edu.cn}

\author{Yi Li}
\affiliation{%
  \institution{Nanyang Technological University}
  \city{Singapore}
  \country{Singapore}
}
\email{yi\_li@ntu.edu.sg}

\begin{document}

\begin{abstract}

As Large Language Model (LLM) agents increasingly automate offensive operations across the cyber kill chain, their efficacy in complex local post-exploitation tasks remains inadequately quantified. Among these, Linux privilege escalation is a key step between initial access and full system compromise. However, existing evaluations for this task are limited by small sample sizes ($<$15~scenarios), lacking the scale to compare model capabilities under executable verification. To address this, we present \textsc{PrivEscalate}, a large-scale benchmark for Linux privilege escalation, comprising 531~Dockerized scenarios spanning 14~distinct sub-categories. To measure sensitivity to environmental distractors, we additionally derive 329~parameterized variant scenarios so that each model's demonstrated successes can be retested under matched perturbations.

Evaluating six LLMs across three agent architectures reveals: (i)~Model capability is heterogeneous across vulnerability classes, with no single model dominating across the high-prevalence classes, motivating multi-dimensional risk assessments. (ii)~LLM successes are sensitive to environmental perturbation, so configuration rotation can disrupt some exploit attempts but does not eliminate the measured risk. (iii)~Agent architectures can materially change success rates and reorder model rankings, though the magnitude is model-dependent. Leveraging these insights, we develop \textsc{PrivEscAgent}, a domain-specialized wrapper that augments a generic ReAct agent with deterministic enumeration, category matching, and step planning. \textsc{PrivEscAgent} improves over prior Linux privilege-escalation agent baselines without underlying LLM modifications. We release \textsc{PrivEscalate} as an open-source, Dockerized measurement instrument supporting both LLM agent evaluation and broader Linux privilege escalation research, including defensive tool validation and red-team training.

\end{abstract}

\maketitle

\section{Introduction}
\label{sec:introduction}

Large language model (LLM)-based autonomous agents are increasingly deployed in offensive cybersecurity~\cite{Malatji2024ArtificialI, Kim2021CyberAttackSM, Ruengsurat2025HumanintheLoopFM}, spanning vulnerability discovery~\cite{fang2024llm_2, lekssays2025llmxcpg}, penetration testing~\cite{deng2024pentestgpt, shen2025pentestagent}, and exploit generation~\cite{wen2025cvebench}. These tasks span the \emph{cyber kill chain}, the canonical sequence of attacker stages: \emph{reconnaissance} gathers target intelligence, \emph{initial access} obtains a constrained foothold, \emph{exploitation} triggers a vulnerability to gain code execution, \emph{privilege escalation}~\cite{Avllazagaj2024SCAVYAD} elevates the foothold to higher system privileges, \emph{lateral movement} expands the compromise across hosts, and \emph{actions on objectives} achieves the attacker's final goal such as data exfiltration. Among these stages, privilege escalation is particularly consequential: it converts a constrained user-context foothold into root-level control over the host, unlocking the rest of the kill chain.

Reproducible benchmarks with executable environments and ground-truth verification have matured for several kill-chain stages~\cite{Kazimierczak2024ImpactOA} by drawing on public corpora: Cybench~\cite{zhang2025cybench} and NYU CTF Bench~\cite{shao2024nyuctf} from CTF write-ups, AutoPenBench~\cite{gioacchini2025autopenbench} from pentest workflows, and CVE-Bench~\cite{wen2025cvebench} from CVE databases. However, privilege-escalation evaluation splits between unreproducible online playgrounds and a single offline benchmark, namely, hackingBuddyGPT~\cite{happe2026llms}, limited to 13 manually constructed scenarios within a narrow slice of the tactic space, while subsequent work such as Perses~\cite{cichon2025perses} adds only a handful more in a few sub-categories. The existing benchmarks suffer from three key limitations: \emph{reproducibility} is undermined by online playgrounds that drift across studies, \emph{scale and coverage} are insufficient because the 13-scenario offline corpus lacks a fine-grained taxonomy, and sensitivity to \emph{environmental distractors} is untested because existing scenarios place vulnerabilities in clean environments.

To address these limitations, we construct \textsc{PrivEscalate}, a measurement framework comprising 531~audited Docker-based Linux privilege escalation scenarios. For \textbf{coverage}, we systematically enumerate a 14-subcategory privesc taxonomy with standardized scenario specifications and populate every sub-category. The final distribution reflects the availability of reproducible, executable privilege-escalation material: Sudo and SUID/SGID dominate over service-heavy or infrastructure-heavy vectors. We report both aggregate and per-sub-category results to separate common-case trends from class-specific behavior. For \textbf{scale}, we design a multi-agent construction pipeline that integrates vulnerability ingestion, scenario generation, and executable verification, achieving a 48\% end-to-end success rate with a mean recorded generation-and-verification cost of \$0.76 for verified scenarios. After expert audit and refinement, the final corpus contains 523 retained scenarios from the automated pipeline and seed set, plus eight manually authored scenarios that fill coverage gaps. For \textbf{validity}, we augment the benchmark with 329 parameterized variants generated by perturbing environmental elements of selected original scenarios, and use per-model success retention as the primary metric for this perturbation analysis.

Based on \textsc{PrivEscalate}, we evaluate six LLMs across three agent architectures, yielding several defender-actionable findings. (1)~\emph{No single model dominates}: the reasoning-augmented model leads overall and in one dominant vulnerability class, while the strongest non-reasoning model leads in another high-prevalence class, so threat assessment must span multiple dimensions. (2)~\emph{Perturbation sensitivity varies sharply across models}: per-model success retention under environmental perturbation ranges from 59.0\% to 78.2\%, so configuration rotation alone cannot fully address the measured risk. (3)~\emph{Agent architecture changes capability beyond raw model choice}: structural design choices can yield larger performance gains than switching the base LLM, but both the effect size and the ordering among baseline agent frameworks are model-dependent. Motivated by these findings, we develop \textbf{\textsc{PrivEscAgent}}, a domain-specialized agent that incorporates deterministic enumeration, category matching, and stepwise planning. \textsc{PrivEscAgent} improves the SR of the strongest non-reasoning model by about 34~percentage points and raises the weakest model to the level of the strongest wintermute baseline, without modifying the underlying LLM. The measured per-success costs show that full-corpus automated evaluation is operationally feasible under our experimental setup.

\noindent In summary, we make the following contributions:
\begin{itemize}[leftmargin=*, topsep=0pt, itemsep=2pt]
    \item \textbf{\textsc{PrivEscalate}: a measurement-grade benchmark for LLM-automated Linux privilege escalation.} We construct 531 audited Docker-based scenarios across a 14-subcategory privesc taxonomy, paired with 329 parameterized variants for perturbation testing. Scenarios are produced by a multi-agent construction pipeline that attains 48\% end-to-end success with a mean recorded generation-and-verification cost of \$0.76 for verified scenarios.

    \item \textbf{An empirical measurement of LLM offensive capability across six LLMs and three agent architectures.} Our findings expose three patterns: model strengths are heterogeneous across vulnerability classes; sensitivity to environmental perturbation varies sharply across models; and agent architecture can materially change success rates and model rankings.

    \item \textbf{\textsc{PrivEscAgent}: a domain-specialized agent for privilege escalation.} We design \textsc{PrivEscAgent} to incorporate deterministic enumeration, category matching, and stepwise planning into the agent loop. It raises SR for all six evaluated models and brings the weakest model to the level of the strongest wintermute baseline, without modifying the underlying LLM.
\end{itemize}

\section{Background}
\label{sec:background}

\subsection{Privilege Escalation in the Cyber Kill Chain}
\label{sec:bg-privesc}

Offensive cyber operations follow a multi-stage \emph{kill chain}~\cite{hutchins2011intelligence, Kouremetis2025OCCULTEL}: Reconnaissance~\cite{Temara2023MaximizingPT}, Initial Access, Privilege Escalation~\cite{Li2025PrivilegeED}, and Post-Exploitation~\cite{Ailabouni2025FGRCAKP}. Initial access (via phishing~\cite{Ahmad2025AcrossTS}, web exploitation~\cite{Priyanka2020WebAV}, credential theft~\cite{Narayanan2025HumanCentricCM}, etc.) typically yields a low-privilege foothold, which is then escalated to unlock post-exploitation objectives requiring higher privileges (lateral movement, data exfiltration, persistence). Privilege escalation, the process of elevating an attacker's access from the initial foothold to higher privileges, is therefore the decisive stage separating constrained attacker impact from full system compromise.

On Linux, a low-privilege user may escalate to higher privileges (e.g., \texttt{root}) by exploiting system misconfigurations, environment hijacking, credential leaks, kernel vulnerabilities, etc. Following standard pentesting methodology~\cite{Hilario2024GenerativeAF, Ginige2025AutoPentesterAL, Zaydi2025GAIDrivenOC}, exploitation proceeds in three phases: enumeration of system state, identification of exploitable weaknesses, and execution of the exploit chain. These phases require integrating heterogeneous signals from the system state and reasoning over multi-step exploit chains, making automation particularly challenging.

\begin{table*}[h]
\centering
\caption{PrivEscalate vulnerability taxonomy.}
\label{tab:taxonomy}
\small
\setlength{\tabcolsep}{4pt}
\renewcommand{\arraystretch}{1.25}
\begin{tabular}{@{}lllp{9.8cm}@{}}
\toprule
\textbf{Class} & \textbf{Sub-category} & \textbf{ATT\&CK} & \textbf{Description} \\
\midrule
\multirow{6}{*}{\textit{Misconfig.}}
  & SUID/SGID abuse & T1548.001 & Shell escape or arbitrary file read through a binary whose setuid-root bit allows root execution \\
  & Sudo misconfiguration & T1548.003 & Password-less or wildcard-permissive sudo rule granting execution of a shell-escape binary \\
  & Capabilities abuse & T1068 & Fine-grained Linux file capability (e.g., calling setuid) attached to a non-root binary \\
  & Polkit misconfiguration & T1548 & Overly permissive polkit rule that grants a low-privilege user unconditional authorization to invoke a privileged action \\
  & D-Bus misconfiguration & T1068 & System D-Bus service that exposes a privileged method without verifying the caller's identity \\
  & Weak file permissions & T1222.002 & World-writable system password or shadow file allowing the attacker to add a UID-0 entry or clear the root password \\
\midrule
\multirow{2}{*}{\textit{Environ.}}
  & PATH hijacking & T1574.007 & Attacker-writable directory placed earlier in the executable search path than system directories \\
  & LD\_PRELOAD hijack & T1574.006 & Privileged process loads attacker-controlled shared library via a preserved dynamic-linker variable \\
\midrule
\multirow{2}{*}{\textit{Sched.}}
  & Cron job exploitation & T1053.003 & Writable script or weakly owned directory invoked by a root-scheduled cron entry \\
  & Systemd service & T1543.002 & Writable systemd service unit executed by root at load \\
\midrule
\multirow{3}{*}{\textit{Cred.}}
  & Password disclosure & T1552.001/.003 & Plaintext credentials recoverable from configuration files, shell history, or environment variables, then reused to authenticate as a privileged local account \\
  & SSH key injection & T1098.004 & Writable \texttt{authorized\_keys} file attached to a privileged account \\
  & DB credential privesc & T1078.003 & Database account whose password is reused by a privileged local OS account, allowing the attacker to authenticate as that account after credential recovery \\
\midrule
\multirow{1}{*}{\textit{Cont.}}
  & Docker/container escape & T1611 & Over-broad container runtime access (group membership, host mount, or accessible socket) \\
\bottomrule
\end{tabular}
\end{table*}

\subsection{LLM-Based Security Agents}
\label{sec:bg-agents}

The enumeration and multi-step reasoning demands of privilege escalation align with recent advances in LLM-based agents~\cite{Zhang2024AgentSB, Lin2025SEAgentST}. LLMs trained on public code and text corpora encode substantial security-relevant knowledge, including vulnerability databases, exploit write-ups, and pentesting curricula. When paired with tool-invocation interfaces (shell execution, function calling), these LLMs form \emph{LLM agents}: autonomous systems that iteratively observe, reason, and act in real environments.

A widely adopted paradigm is ReAct~\cite{yao2023react}, an iterative loop in which the agent \emph{observes} command output, \emph{reasons} about its security implications, and \emph{acts} by issuing the next shell command, repeating until the objective is achieved or a bounded step budget is exhausted. Variants such as Planner-Summarizer agents~\cite{muzsai2024hacksynth} add structured memory on top of the ReAct loop. This fundamentally differs from traditional enumeration scripts that execute a static checklist: LLM agents can adaptively interpret novel outputs, chain multi-step exploits across different system components, and recover from failed attempts by reasoning about error messages. The combination of broad security knowledge encoded in pre-training data and the ability to execute arbitrary commands makes these agents a qualitatively new class of offensive automation, raising significant concerns about their potential for autonomous exploitation in real-world systems. While privilege escalation is the decisive stage of the cyber kill chain and a particularly challenging benchmark for evaluating LLM agents' offensive capabilities on Linux, existing evaluations remain limited in scope and realism. As a result, the true effectiveness of LLM agents in complex, multi-step privilege escalation scenarios, and the extent to which their capabilities can be systematically evaluated and further enhanced, remain largely unknown.

\subsection{Threat Model}
\label{sec:threat-model}

We consider a post-initial-access adversary who holds an unprivileged interactive shell on a Linux host and aims to escalate to root privileges. The adversary acts entirely through an autonomous LLM agent that issues arbitrary shell commands~\cite{Andrew2022MappingLS}, without GUI access or human-in-the-loop assistance; we exclude cross-host lateral movement, kernel-CVE exploitation, and post-root objectives such as data exfiltration and persistence. The adversary may exploit any escalation path the local Linux configuration exposes, regardless of vulnerability class. Crucially, the agent operates under zero-knowledge conditions: it receives no hints about the vulnerability category, exploitable binary, or exploitation path, and must discover them from interactions.

\section{Benchmark Design and Construction}
\label{sec:benchmark}

\textsc{PrivEscalate} addresses the three measurement limitations introduced in \Cref{sec:introduction}: reproducibility, scale and coverage, and sensitivity to environmental distractors. Specifically, we enumerate Linux-relevant privilege-escalation sub-techniques, filter for Docker feasibility, and populate all 14 retained sub-categories with executable scenarios to broaden coverage (\Cref{sec:taxonomy}); use an automated multi-agent pipeline producing Dockerized scenarios with executable verification to support reproducibility and scalable construction (\Cref{sec:privescgen}), refined by expert audit (\Cref{sec:audit}); and generate parameterized variant scenarios that inject environmental noise to measure perturbation sensitivity (\Cref{sec:variant-robustness}).

\subsection{Taxonomy}
\label{sec:taxonomy}

PrivEscalate organizes vulnerability scenarios using a taxonomy derived systematically from the MITRE ATT\&CK framework~\cite{mitre2024attack} through a four-step process:

\noindent\ding{172}\ \textbf{Enumeration.} We enumerated Linux-relevant ATT\&CK techniques and sub-techniques that can realize local privilege escalation, starting from tactic TA0004 and retaining cross-tactic techniques when their documented use elevates local privileges.

\noindent\ding{173}\ \textbf{Feasibility filtering.} We excluded sub-techniques that cannot be reliably reproduced in Docker containers; \Cref{app:excluded-categories} details the excluded categories and per-category rationale.

\noindent\ding{174}\ \textbf{Cross-validation.} We verified each remaining category against two community-maintained technique databases, GTFOBins~\cite{gtfobins} and Exploit-DB~\cite{exploitdb}, to confirm each category has documented real-world instances.

\noindent\ding{175}\ \textbf{Continuity check.} To preserve comparability with prior work, we verified that every scenario in the hackingBuddyGPT benchmark maps into a sub-category of our taxonomy.

\noindent This process yielded 14~sub-categories, which we group by the nature of the vulnerability into 5 classes for analytical convenience: \emph{Misconfiguration} (Misconfig.) covers permission and identity-setting mistakes, \emph{Environment} (Environ.) covers process-environment resolution abuse, \emph{Scheduled Tasks} (Sched.) covers time-triggered execution, \emph{Credentials} (Cred.) covers credential storage and reuse, and \emph{Container} (Cont.) covers container-boundary violations. Each sub-category is mapped to a specific ATT\&CK technique and a concise exploitation-pattern description. \Cref{tab:taxonomy} presents the complete taxonomy.

\subsection{Automated Construction Pipeline}
\label{sec:privescgen}

\begin{figure}[h]
    \centering
    \includegraphics[width=\linewidth]{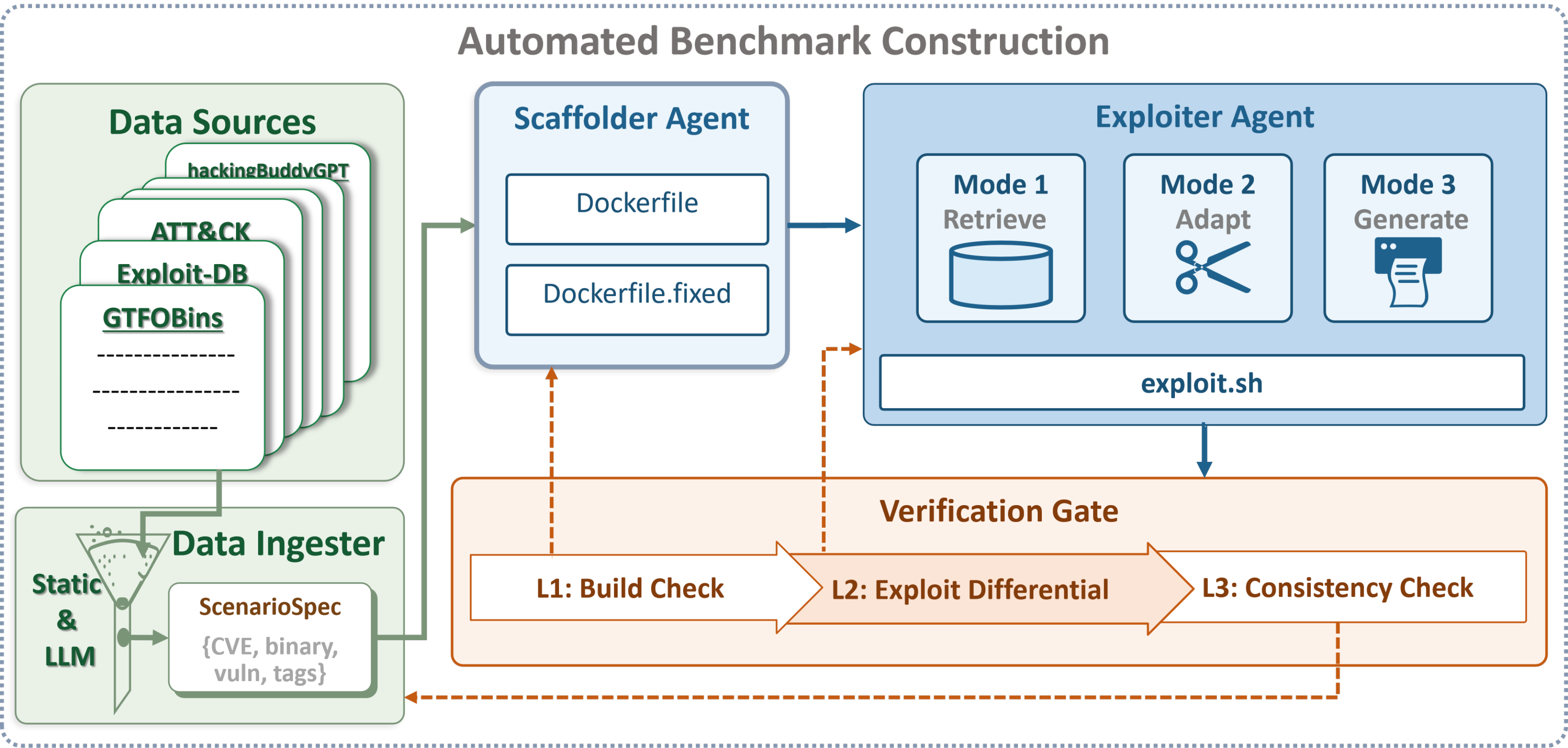}
    \caption{PrivEscalate framework and construction pipeline overview.}
    \Description{Flow diagram showing vulnerability ingestion, scenario generation, verification, and benchmark evaluation stages.}
    \label{fig:overview}
\end{figure}

\Cref{fig:overview} presents the overall architecture: starting from structured exploit databases, a multi-agent pipeline ingests candidate vulnerabilities, scaffolds Dockerized environments, produces ground-truth exploits, and verifies each scenario through differential testing.

\subsubsection{Data Sources}

The pipeline consumes two inputs: (i)~GTFOBins, a public catalog of Unix binaries with documented abuse techniques, from which we take SUID, sudo, and capabilities categories relevant to privilege escalation; and (ii)~Exploit-DB, a public archive of exploit scripts for publicly-disclosed software vulnerabilities, from which we take Linux local privilege-escalation entries.

\subsubsection{DataIngester}
\label{sec:dataingester}

Most publicly documented privilege escalation entries, including those in GTFOBins and Exploit-DB, come as free-form prose (vulnerability narratives, CVE writeups, and partial PoC fragments) rather than machine-readable scenario specifications, which makes automated reproduction brittle. To bridge this gap, we design the \textbf{DataIngester} to convert each raw entry into a uniform \texttt{ScenarioSpec} template, shown in \Cref{fig:scenariospec}, that captures the minimum information needed to materialize and verify a scenario. Every subsequent module (Scaffolder, Exploiter, and Verifier) then analyzes and processes this template, so each stage consumes a single structured input instead of re-parsing heterogeneous prose.

\begin{figure}[h]
\begin{lstlisting}[language=json]
{
  "template_id":      "...", // unique scenario id
  "category":         "...", // taxonomy class
  "attack_technique": "...", // MITRE ATT&CK TA0004 id
  "description":      "...", // vuln summary+PoC source
  "docker_run_args":  [...], // extra docker run flags
  "params":           {...}, // runtime params(user/pass)
  "status":           "..."  // runtime-updated outcome
}
\end{lstlisting}
\caption{The \texttt{ScenarioSpec} schema.}
\Description{JSON schema listing scenario identifier, category, ATT\&CK technique, description, Docker options, parameters, and status.}
\label{fig:scenariospec}
\end{figure}


\noindent The DataIngester handles each source according to its format. \textbf{Structured sources} (e.g., GTFOBins) provide an index that clearly labels the fields and metadata per entry; the DataIngester reads the index and maps it into a \texttt{ScenarioSpec} via deterministic static-table lookups, reducing parser ambiguity for these entries. \textbf{Description-based sources} (e.g., Exploit-DB) provide only unstructured descriptions; for each entry, the DataIngester first runs an LLM-based feasibility classifier that discards it if not reproducible in a Docker container, then uses an LLM parser to extract the retained entry into a \texttt{ScenarioSpec}.

The DataIngester additionally runs two cross-source operations. (1)~\emph{Reference feedback loop}: every verified scenario (its \texttt{ScenarioSpec} together with its Dockerfiles) is added to a category-indexed pool; for each new candidate, the DataIngester picks a same-category entry from this pool and attaches it as a one-shot example for the Scaffolder, providing increasingly many same-category exemplars as construction proceeds. (2)~\emph{Full-coverage discovery}: the DataIngester enumerates every exploitable binary and every feasibility-filtered Exploit-DB entry without sampling, and skips entries already ingested in prior runs.

\subsubsection{Scaffolder}
\label{sec:scaffolder}

The \textbf{Scaffolder} materializes each \texttt{ScenarioSpec} into a containerized exploit environment, producing two Dockerfiles that differ only in whether the target vulnerability is injected or removed. It takes as input the target \texttt{ScenarioSpec} along with a one-shot (\texttt{ScenarioSpec}, Dockerfiles) pair from a previously-verified scenario.

The Scaffolder combines a static hardening template with LLM-driven Dockerfile generation. First, \textbf{base hardening}: before the vulnerability is injected, it applies a hardening step that preserves authentication-gated SUID/SGID binaries and locks down ambient escalation paths (stale cron configurations, extraneous sudoers rules), so the injected vulnerability becomes the sole escalation path while the environment remains realistic. Next, \textbf{LLM-driven generation}: starting from a minimal base image and guided by the one-shot reference, the LLM emits in a single call a vulnerable Dockerfile (injecting the target vulnerability) and a matched fixed Dockerfile (the same environment with the vulnerability removed); the generation also explicitly resolves and installs required dependencies (target binary, language runtimes, kernel-utility or language-specific modules) based on the vulnerability description and exploit approach. On a build failure, the LLM diagnoses the error log and attempts to produce a corrected Dockerfile.

\subsubsection{Exploiter}
\label{sec:exploiter}

The \textbf{Exploiter} produces a ground-truth exploit script for the vulnerable environment. It takes as input the target \texttt{ScenarioSpec} together with the Scaffolder's vulnerable Dockerfile.

The Exploiter operates in three progressive modes selected by a deterministic rule, drawing on an exploit knowledge base as its reference library. \textbf{Retrieval} reuses a reference exploit directly when an exact match exists for the binary and exploit type. \textbf{Adaptation} takes a same-category reference exploit and has the LLM adjust parameter differences, used when no exact match is found for the binary and exploit type. \textbf{Generation}, the fallback when neither preceding mode applies, has the LLM construct the exploit from scratch using the vulnerability description and same-category references as context.

\subsubsection{Verifier and Feedback Loop}
\label{sec:verifier}

The \textbf{Verifier} is a script-driven module that performs triple verification on every generated scenario, designed together with Scaffolder's base hardening to check that the declared exploit path is present while common unintended paths are suppressed:
\begin{enumerate}
    \item \textbf{Build check.} Confirms that both the vulnerable and fixed Dockerfiles build and run successfully so that the paired environments are reproducible.
    \item \textbf{Exploit differential check.} Requires the exploit script to elevate to root on the vulnerable container but \emph{fail} on the fixed container, confirming that the exploit targets the intended vulnerability and that the fix effectively removes it.
    \item \textbf{Consistency check.} Runs a suite of automated enumeration probes on the fixed container, covering SUID/SGID binaries, sudo rules, cron jobs, Linux capabilities, writable sensitive files, and shell users, and flags non-standard vectors for review to rule out unintended privilege-escalation paths.
\end{enumerate}

When verification fails, a \textbf{Manager} module orchestrates an LLM-driven feedback cycle, bounded by a retry limit per scenario. The Verifier produces a structured diagnosis identifying whether the root cause lies in the Dockerfile, the exploit, or both, and the Manager dispatches a targeted fix to the responsible sub-agent (Scaffolder or Exploiter). The Verifier then re-runs the triple check, closing the verify--diagnose--fix--re-verify loop.

\subsection{Construction Statistics}
\label{sec:construction-stats}

\paragraph{Setup}
We seed the DataIngester's reference pool with the 13~hackingBuddyGPT scenarios~\cite{happe2026llms}, and for each we author a matched fixed Dockerfile and \texttt{ScenarioSpec} to serve as the Scaffolder's initial one-shot references. The DataIngester then ingests two exploit databases, GTFOBins (478~binaries, commit \texttt{c922862e}) and Exploit-DB (527~entries, commit \texttt{a0b1c92c}), which also form the Exploiter's knowledge base. These sources yield 1{,}067~\emph{pipeline-input templates}: 771 from GTFOBins, emitting one template per exploit type per binary, and 296 from Exploit-DB after feasibility filtering that drops kernel exploits and entries infeasible to containerize; the full filter chain appears in \Cref{app:pipeline-details}. Each template then runs through the construction pipeline with the feedback cycle capped at 3~retries per scenario. We use \textbf{Claude Opus~4.6} as the backing LLM for all pipeline stages. The recorded construction logs total \$1{,}044 in API cost; among verified scenarios with cost logs, the mean generation-and-verification cost is \$0.76.

\begin{figure}[h]
    \centering
    \includegraphics[width=\linewidth]{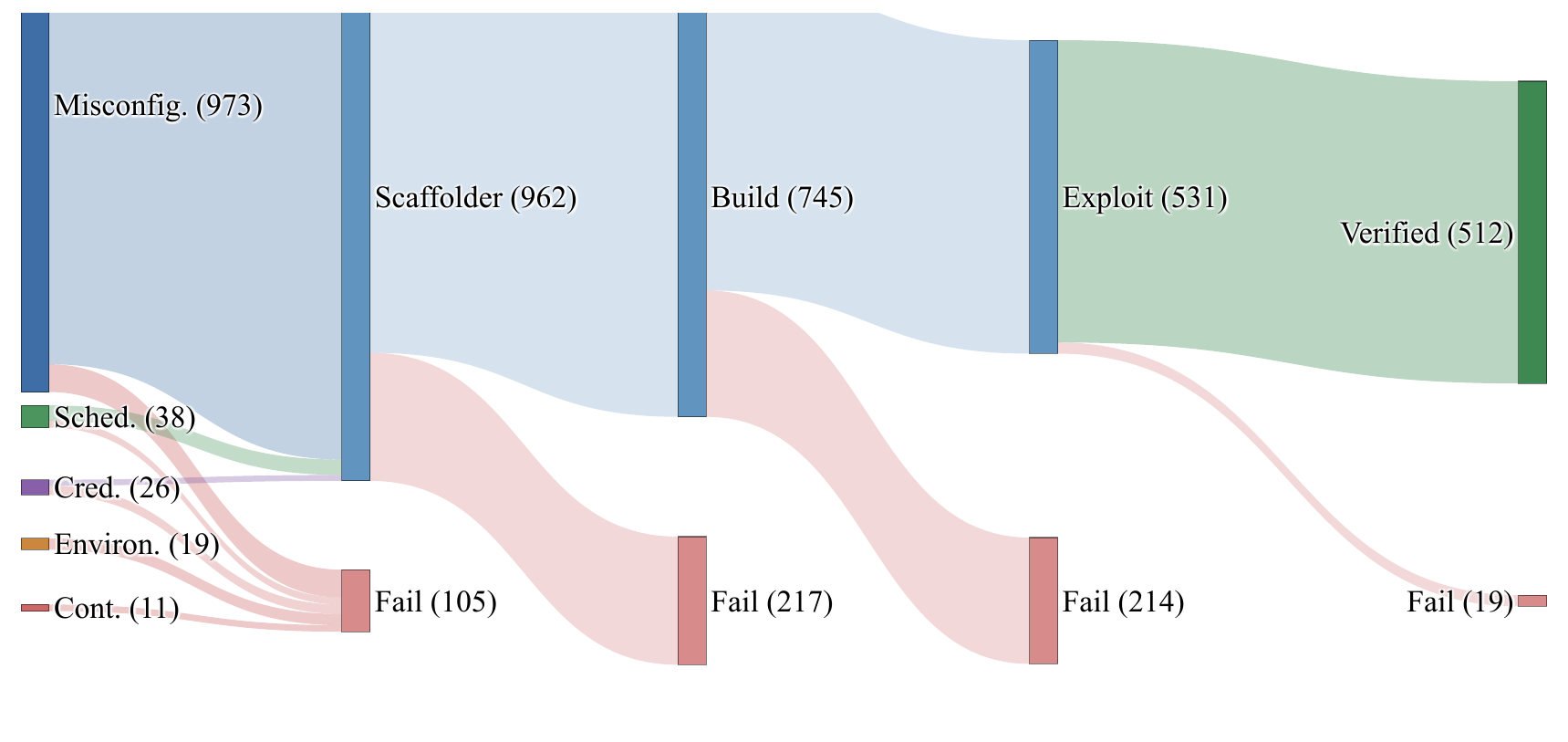}
    \caption{Construction pipeline flow from 1{,}067 candidates to 512 verified scenarios.}
    \Description{Sankey diagram showing candidate templates passing through scaffolding, build verification, exploit verification, and final verified scenarios.}
    \label{fig:pipeline-funnel}
\end{figure}

\paragraph{Results}
The pipeline processed 1{,}067 candidate templates and verified 512 of them (48.0\%; \Cref{fig:pipeline-funnel}). Of the 555 failed cases, 105 (18.9\%) failed at the Scaffolder stage (LLM unable to generate a valid Dockerfile), 217 (39.1\%) failed the Verifier's build check (Dockerfile build failure), 214 (38.6\%) failed the Verifier's exploit check on the vulnerable image (exploit did not elevate to root), and 19 (3.4\%) failed on the fixed image (fix did not block the exploit). The most common failure modes are (1)~LLM-generated exploit scripts that do not allocate a pseudo-terminal required by interactive target binaries, (2)~LLM-generated Dockerfiles with missing package dependencies, and (3)~LLM-generated scripts that implement only part of a multi-step exploit (e.g., file read or SUID copy) without reaching a root shell.

\begin{table}[h]
\centering
\caption{Per-category benchmark composition.}
\label{tab:construction}
\scriptsize
\setlength{\tabcolsep}{3pt}
\begin{tabular}{@{}llrrrrr@{}}
\toprule
\textbf{Class} & \textbf{Category} & \textbf{Cand.} & \textbf{Verif.} & \textbf{Rate} & \textbf{hBGPT} & \textbf{Exp.} \\
\midrule
\textit{Misconfig.} & SUID/SGID    & 462 & 204 & 44\% & 1 & 0 \\
                    & Sudo         & 479 & 290 & 61\% & 3 & 0 \\
                    & Capabilities & 10  & 8   & 80\% & 0 & 0 \\
                    & Polkit       & 9   & 0   & 0\%  & 0 & 1 \\
                    & D-Bus        & 13  & 0   & 0\%  & 0 & 1 \\
                    & Weak perms   & 0   & 0   & ---  & 0 & 2 \\
\textit{Environ.}   & PATH hijack  & 10  & 0   & 0\%  & 0 & 1 \\
                    & LD\_PRELOAD  & 9   & 0   & 0\%  & 0 & 1 \\
\textit{Sched.}     & Cron         & 30  & 8   & 27\% & 2 & 0 \\
                    & Systemd      & 8   & 0   & 0\%  & 0 & 1 \\
\textit{Cred.}      & Password     & 18  & 2   & 11\% & 5 & 0 \\
                    & SSH key      & 1   & 0   & 0\%  & 1 & 0 \\
                    & DB cred.     & 7   & 0   & 0\%  & 0 & 1 \\
\textit{Cont.}      & Docker esc.  & 11  & 0   & 0\%  & 1 & 0 \\
\midrule
\multicolumn{2}{l}{\textbf{Total}} & \textbf{1{,}067} & \textbf{512} & \textbf{48\%} & \textbf{13} & \textbf{8} \\
\bottomrule
\end{tabular}
\end{table}

\Cref{tab:construction} breaks down these outcomes by sub-category. Three structured misconfiguration categories pass reliably: Capabilities (8/10, 80\%), Sudo (290/479, 61\%), and SUID/SGID (204/462, 44\%). Each vulnerability amounts to a single permission-flag change on a target binary, so the Scaffolder's task reduces to generating one deterministic Dockerfile directive with no additional environment setup. Eight categories fail at 0\%, splitting into two groups by root cause. Five are \emph{interaction-heavy} (Polkit, D-Bus, systemd, LD\_PRELOAD, PATH hijack): they require orchestrating service units, writable policy files, or environment-variable inheritance chains that span multiple container layers, which the Scaffolder routinely misspecifies. The other three (Docker escape, SSH key, DB credential) are \emph{infrastructure-constrained}: Docker escapes need privileged containers or kernel-level semantics incompatible with our shared-kernel Docker setup, while SSH-key and DB-credential exploits depend on specific service versions or authentication flows that exceed the Scaffolder's single-container reproduction scope.

To close the pipeline's zero-coverage sub-categories, we hand-author 8 expert scenarios under the same SSH/Docker interface and differential-verification protocol as pipeline scenarios (\emph{Exp.}\ column of \Cref{tab:construction}): 6 target the five interaction-heavy categories and DB credential; 2 target Weak-file-permissions, absent from both data sources. Before audit, construction yields \textbf{533 candidate scenarios}: 512 verified pipeline outputs, 13 hackingBuddyGPT seed scenarios, and 8 expert scenarios.

\subsection{Benchmark Quality Audit}
\label{sec:audit}

Beyond the automated triple verification, we manually audit a stratified sample of scenarios on two quality dimensions: \emph{Dockerfile correctness}, whether the Docker environment matches the declared vulnerability spec, and \emph{exploit validity}, whether the ground-truth exploit reaches root via the declared target vulnerability rather than a secondary path. The audit covers 82 scenarios from the 525 scenarios produced by the automated pipeline or seed set (512 pipeline outputs plus 13 hackingBuddyGPT seeds), providing $\pm 10\%$ margin of error at 95\% confidence on the true rate of issues.

A security expert with five years of offensive-security experience trained two auditors, a security-background PhD student and a research engineer, on the audit rubric. The two auditors then independently graded each sampled scenario on the two quality dimensions and also assigned an overall scenario verdict. For each row in \Cref{tab:audit-verdicts}, a scenario falls into one of three buckets: \emph{unanimous-approve}, \emph{unanimous-flag}, or \emph{disagreement}. The first two rows report issue localization by dimension; the final row reports an independently assigned scenario-level disposition.

\begin{table}[h]
\centering
\caption{Audit verdicts by dimension and overall scenario disposition.}
\label{tab:audit-verdicts}
\resizebox{\columnwidth}{!}{%
\begin{tabular}{@{}lccc@{}}
\toprule
\textbf{Audit row} & \textbf{Unanimous-approve} & \textbf{Unanimous-flag} & \textbf{Disagreement} \\
\midrule
Dockerfile correctness & 71 (86.6\%) & 5 (6.1\%) & 6 (7.3\%) \\
Exploit validity       & 72 (87.8\%) & 3 (3.7\%) & 7 (8.5\%) \\
\midrule
Overall scenario verdict & 68 (82.9\%) & 10 (12.2\%) & 4 (4.9\%) \\
\bottomrule
\end{tabular}}
\end{table}

At the scenario level, 10 cases were unanimously flagged as requiring review and 4 elicited an overall disagreement. After the expert adjudicated each flagged or disputed case together with the auditors, the 4 overall disagreements were retained unchanged (the expert determined none reflected substantive issues), 5 of the 10 unanimous flags were converted into actionable adjustments, and the remaining 5 unanimous flags were retained after re-review; this leaves 79 of the 82 sampled scenarios (96.3\%) with verified Dockerfile correctness and 80 (97.6\%) with verified exploit validity. Independently, the expert conducted a broader audit across the full 525-scenario automated-pipeline and seed-set pool, identifying 6 further scenarios in which the ground-truth exploit relied on a secondary path (e.g., writable \texttt{/etc/passwd}) rather than the declared target tool; all 6 were reclassified. Combined, the audit yielded 11 actionable metadata or taxonomy adjustments.

After adjudication, the audit triggered two classes of adjustment: 2 scenarios were removed because their declared exploit primitives were inoperative on modern Linux distributions, and 9 had their sub-category label corrected within the taxonomy. The resulting 531-scenario benchmark contains 523 audit-retained, 2 weak-permissions, and 6 expert-authored scenarios; we name this final evaluation set \textsc{PrivEscalate}. These 531 scenarios form the original corpus; for perturbation testing we additionally generate 329 perturbed \emph{variants} by altering environmental elements of selected original scenarios (\Cref{sec:variant-robustness}).

\section{Measurement Study}
\label{sec:measurement}

\subsection{Research Questions}
\label{sec:rqs}

We formulate four research questions on LLM-automated privilege escalation under zero-knowledge conditions:
\begin{itemize}[leftmargin=*, itemsep=2pt, topsep=2pt]
    \item \textbf{RQ1 (Model Threat Ranking)}: How do current LLMs perform on automated Linux privilege escalation?
    \item \textbf{RQ2 (Variant Perturbation)}: How stable is model success under environment perturbations that preserve the exploit primitive?
    \item \textbf{RQ3 (Agent Architecture Impact)}: How does agent architecture affect the automated escalation threat, and does the effect depend on model class?
    \item \textbf{RQ4 (Cost and Efficiency)}: What is the per-attempt and per-success cost, and how many steps do successful escalations consume, i.e., how narrow is the defender's detection window?
\end{itemize}

\subsection{Setup}
\label{sec:experiment}

\noindent\textbf{Models.} We evaluate six LLMs spanning frontier, cost-efficiency, proprietary API, and open-family baselines: GPT-5.4 (reasoning-augmented frontier; its inference mode lets us assess whether explicit reasoning affects escalation performance), Claude Sonnet~4.6 (non-reasoning frontier), GPT-4.1 (OpenAI prior-generation baseline), Claude Haiku~4.5 (cost-efficiency baseline), DeepSeek~v3.2 (cost-effective open-family baseline), and Qwen-Plus (multilingual API-served baseline; snapshot \texttt{qwen-plus-2025-12-01}). None of these models was used during benchmark construction, reducing benchmark-in-the-loop contamination from our pipeline.

\noindent\textbf{Agent Frameworks.} RQ1 and RQ2 use \textbf{hackingBuddyGPT wintermute}~\cite{happe2026llms}, a ReAct-based framework, as the baseline agent. RQ3 and RQ4 compare wintermute with \textbf{HackSynth}~\cite{muzsai2024hacksynth}, a Planner-Summarizer architecture representing a planning-oriented paradigm distinct from ReAct.

\noindent\textbf{Configuration.} Experiments run under \textbf{zero-knowledge} conditions: the agent starts from a low-privilege SSH foothold and must achieve \texttt{uid=0(root)} through arbitrary shell commands, receiving no hints about the vulnerability category or exploitation method. Parameters: \textbf{max 20~steps} per attempt, \textbf{temperature 0}, SSH interface. The 20-step budget follows hackingBuddyGPT and Perses~\cite{happe2026llms,cichon2025perses}, enabling direct comparison with prior Linux privilege-escalation agent studies. It is a cap rather than a fixed trace length: episodes terminate immediately after root is reached.

\noindent\textbf{Statistical Analysis.} Our primary metric is \textbf{Success Rate (SR)}, the fraction of scenarios where the agent achieves \texttt{uid=0(root)} within the step budget; we also report per-scenario API cost. For paired comparisons on the same scenario set---the original--variant pairs in RQ2, the agent comparisons in RQ3, and the ablation configurations---we use McNemar's test on discordant outcomes. For RQ2, we report success retention as the primary perturbation metric: among scenarios a model solved on the original corpus, the retention rate is the fraction it also solves after perturbation. We also report Pearson correlation over paired outcomes. Each tuple in the primary evaluation is evaluated once under the fixed configuration above; a separate one-run wintermute repeatability check is summarized in \Cref{sec:limitations}.

\subsection{RQ1: Model Threat Ranking}
\label{sec:results}

\begin{figure}[h]
    \centering
    \includegraphics[width=\columnwidth]{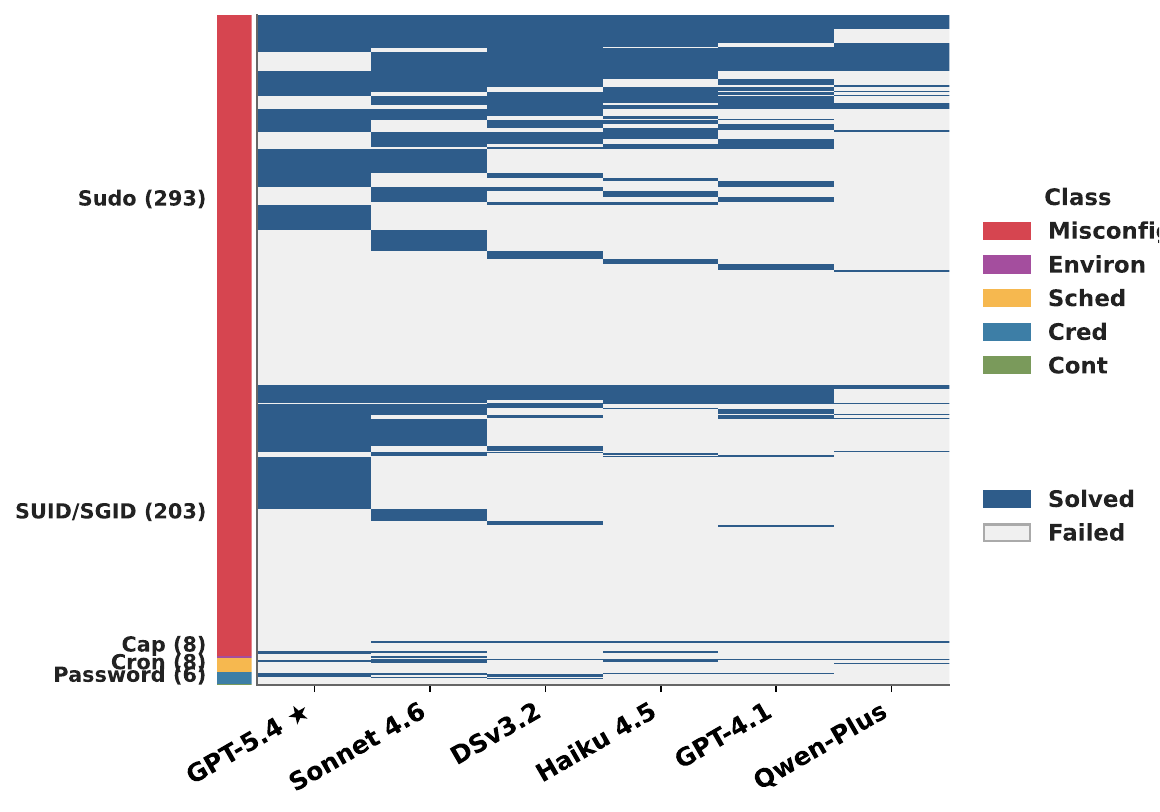}
    \caption{Six-model solution map (RQ1).}
    \Description{Heatmap indicating which benchmark scenarios are solved by each of the six evaluated models under the wintermute baseline.}
    \label{fig:rq1-solution-map}
\end{figure}

We evaluate the six models on the 531 original scenarios of \textsc{PrivEscalate}. \Cref{tab:results-main} reports SR by vulnerability class, \Cref{tab:per-sub-cat-sr} gives the per-sub-category breakdown, and \Cref{fig:rq1-solution-map} visualizes the per-scenario solution map. The aggregate results primarily reflect public, reproducible privilege-escalation material, where Sudo and SUID/SGID cases are common; the per-sub-category breakdown shows how model behavior varies beyond those dominant classes.

\begin{table}[h]
\centering
\caption{SR (\%) by model and vulnerability class.}
\label{tab:results-main}
\begin{tabular}{@{}lccccc@{}}
\toprule
\textbf{Model} & \textbf{SUID} & \textbf{Sudo} & \textbf{Cap.} & \textbf{Other} & \textbf{Overall} \\
\midrule
\textbf{GPT-5.4} & \textbf{45.8} & 39.9 & 0.0 & 25.9 & \textbf{40.9} \\
Claude Sonnet~4.6 & 27.6 & \textbf{46.1} & \textbf{12.5} & \textbf{29.6} & 37.7 \\
DeepSeek~v3.2 & 13.3 & 37.5 & \textbf{12.5} & 18.5 & 26.9 \\
GPT-4.1 & 12.3 & 30.4 & \textbf{12.5} & 7.4 & 22.0 \\
Claude Haiku~4.5 & 8.9 & 32.8 & \textbf{12.5} & 14.8 & 22.4 \\
Qwen-Plus & 3.4 & 15.0 & \textbf{12.5} & 7.4 & 10.2 \\
\bottomrule
\end{tabular}
\end{table}

\begin{table*}[t]
\centering
\caption{Per-sub-category SR (\%) under wintermute.}
\label{tab:per-sub-cat-sr}
\scriptsize
\setlength{\tabcolsep}{3pt}
\resizebox{\textwidth}{!}{%
\begin{tabular}{@{}llrrrrrrr@{}}
\toprule
\textbf{Sub-category} & \textbf{ATT\&CK} & \textbf{N} & \textbf{GPT-5.4} & \textbf{Sonnet} & \textbf{DSv3.2} & \textbf{GPT-4.1} & \textbf{Haiku} & \textbf{Qwen+} \\
\midrule
Sudo misconfiguration      & T1548.003      & 293 & 39.9          & \textbf{46.1} & 37.5          & 30.4 & 32.8 & 15.0 \\
SUID/SGID abuse            & T1548.001      & 203 & \textbf{45.8} & 27.6          & 13.3          & 12.3 & 8.9  & 3.4  \\
Capabilities abuse         & T1068          & 8   & 0.0           & 12.5          & 12.5          & 12.5 & 12.5 & 12.5 \\
Cron job exploitation      & T1053.003      & 10  & 10.0          & \textbf{30.0} & 10.0          & 10.0 & 20.0 & 20.0 \\
Password disclosure         & T1552.001/.003 & 7   & 42.9          & 28.6          & 42.9          & 14.3 & 14.3 & 0.0  \\
SSH key injection           & T1098.004      & 1   & 0.0           & \textbf{100.0} & 0.0           & 0.0           & 0.0  & 0.0  \\
Weak file permissions      & T1222.002     & 2   & \textbf{100.0}& 50.0          & 0.0           & 0.0  & 50.0 & 0.0  \\
Docker/container escape    & T1611          & 1   & \textbf{100.0}& 0.0           & \textbf{100.0}& 0.0  & 0.0  & 0.0  \\
LD\_PRELOAD hijack         & T1574.006      & 1   & 0.0           & \textbf{100.0}& 0.0           & 0.0  & 0.0  & 0.0  \\
Polkit misconfiguration    & T1548          & 1   & 0.0           & 0.0           & 0.0           & 0.0  & 0.0  & 0.0  \\
D-Bus misconfiguration     & T1068          & 1   & 0.0           & 0.0           & 0.0           & 0.0  & 0.0  & 0.0  \\
PATH hijacking             & T1574.007      & 1   & 0.0           & 0.0           & 0.0           & 0.0  & 0.0  & 0.0  \\
Systemd service            & T1543.002      & 1   & 0.0           & 0.0           & 0.0           & 0.0  & 0.0  & 0.0  \\
DB credential privesc      & T1078.003      & 1   & 0.0           & 0.0           & 0.0           & 0.0  & 0.0  & 0.0  \\
\midrule
\textbf{Total}             & ---            & 531 & 40.9          & 37.7          & 26.9          & 22.0 & 22.4 & 10.2 \\
\bottomrule
\end{tabular}}
\end{table*}

\noindent\textbf{Results.} \Cref{tab:results-main}, \Cref{tab:per-sub-cat-sr}, and \Cref{fig:rq1-solution-map} reveal four structural patterns. \textbf{(1)~No single model dominates the dominant classes}: GPT-5.4, a \textbf{reasoning-augmented} model, leads overall SR (40.9\%) and SUID/SGID abuse (45.8\%), while Claude Sonnet~4.6 (37.7\% overall) leads Sudo misconfiguration (46.1\%). This split across the two high-prevalence categories shows that the aggregate ranking is not driven by one uniformly strongest model. The strongest/weakest aggregate gap is 4.0$\times$ (40.9\% vs.\ 10.2\%), so model choice meaningfully shifts the threat level. \textbf{(2)~Behavior varies beyond the dominant classes}: the per-sub-category breakdown shows additional model-specific strengths across capabilities, credentials, cron, systemd, and package-management scenarios, reinforcing the need to inspect category-level behavior alongside aggregate SR. \textbf{(3)~Multi-model union expands the threat surface}: the union of scenarios solved by any of the six models is 329/531 (62.0\%), approximately 50\% higher than the best single model's 40.9\%; two models combined (GPT-5.4 and Claude Sonnet~4.6, $298/531 = 56.1\%$) already capture 90\% of the six-model ceiling, while the other four contribute only $+5.8$\,pp. \textbf{(4)~Reasoning expands rather than intersects the threat surface}: GPT-5.4 uniquely solves 63~scenarios (29.0\% of its solved set) and Claude Sonnet uniquely solves 30 (15.0\%), so reasoning-augmented and non-reasoning models cover different subsets even under the same agent framework.

\begin{tcolorbox}[colback=blue!5, colframe=blue!40!black]
\textbf{Finding 1:} Across our evaluated panel, the multi-model union covers substantially more scenarios than any single best model, so single-model assessments can understate measured exposure.
\end{tcolorbox}

\noindent\textbf{Failure mode analysis.} We classify 2{,}336 failed runs across all six models into three modes: \emph{Wrong-method} (failed despite attempting exploitation), \emph{No-attempt} (vulnerability located but no exploit issued), and \emph{Loop} (the identical command issued three or more times in a row). \Cref{tab:failure-modes} shows the distribution of failure modes and the share of enumeration commands across failed runs, revealing three patterns. First, \emph{Wrong-method} dominates failures (81.6\% across six models; 79.5\% across the five non-reasoning models): exploit attempts are issued but fail to reach root, suggesting the bottleneck is method selection (a domain-knowledge problem) rather than recognition of the vulnerability. Second, Qwen-Plus is the clear \emph{Loop} outlier (15.1\% vs.\ $\leq$2.2\% for the other five models), repeating identical commands rather than progressing. Third, compared to non-reasoning models, GPT-5.4 relies far less on enumeration (21.7\% vs. 42.0–72.4\%), indicating more targeted and effective exploitation.
\begin{table}[h]
\centering
\caption{Failure Mode Distribution (\%)}
\label{tab:failure-modes}
\small
\begin{tabular}{@{}lrrrr@{}}
\toprule
\textbf{Model} & \textbf{Wrong-method} & \textbf{No-attempt} & \textbf{Loop} & \textbf{Enum} \\
\midrule
Claude Sonnet~4.6 & 94.9 & 3.3 & 1.8 & 42.0 \\
Claude Haiku~4.5 & 79.9 & 18.0 & 2.2 & 58.4 \\
GPT-4.1 & 89.1 & 9.9 & 1.0 & 69.6 \\
DeepSeek~v3.2 & 79.6 & 18.8 & 1.5 & 65.7 \\
Qwen-Plus & 60.0 & \textbf{24.9} & \textbf{15.1} & 72.4 \\
GPT-5.4 & \textbf{95.2} & 3.5 & 1.3 & \textbf{21.7} \\
\midrule
\textbf{All (W. Avg.)} & \textbf{81.6} & 14.1 & 4.3 & 57.8 \\
\bottomrule
\end{tabular}
\end{table}

\begin{tcolorbox}[colback=blue!5, colframe=blue!40!black]
\textbf{Finding 2:} Among observed failures, choosing an effective exploit method is a larger bottleneck than deciding whether to attempt exploitation.
\end{tcolorbox}

\begin{tcolorbox}[colback=blue!5, colframe=blue!40!black]
\textbf{Finding 3:} Repeated enumeration dominates failed-run command budgets.
\end{tcolorbox}

\noindent\textbf{Strategy-fixation analysis.} We label a failed episode as \emph{fixated} when three or more consecutive commands invoke the same binary, allowing flag and argument variation; this is a relaxation of the exact-string \emph{Loop} mode in \Cref{tab:failure-modes}, so every \emph{Loop} episode is also fixated but the converse does not hold. Fixation is model-dependent: Qwen-Plus 98.5\%, Claude Haiku~4.5 95.6\%, Claude Sonnet~4.6 83.7\%, GPT-4.1 81.4\%, GPT-5.4 67.8\%, DeepSeek~v3.2 44.3\%. Four of the six models fixate in 81 to 99\% of failed episodes, while DeepSeek~v3.2 and GPT-5.4 diversify earlier.

\begin{tcolorbox}[colback=blue!5, colframe=blue!40!black]
\textbf{Finding 4:} Most evaluated models retry the same strategy class on failure, motivating an automatic pivot mechanism to escape this fixation.
\end{tcolorbox}

\subsection{RQ2: Variant Perturbation}
\label{sec:variant-robustness}

Building on RQ1, we test whether model success transfers across surface-level environment changes. A high aggregate SR may depend on environmental cues rather than the underlying exploit primitive; variant testing measures how much configuration rotation disrupts demonstrated successes. We derive the variant set from original scenarios solved by at least one model in the wintermute baseline, yielding 329 unique perturbed variants. Within this set, 154 variants form a shared panel selected from scenarios solved by at least two models and are used for Pearson VR analysis. For retention analysis, we evaluate the matched variant for every RQ1 success of each model, yielding 850 model-variant pairs; the denominator is model-specific and equals the \emph{RQ1 solved} column in \Cref{tab:results-vr}. Each variant preserves the exact vulnerability mechanism while altering the environment context along the following five transformation axes:

\noindent\ding{172}\ \textbf{Credentials}: changed username, password, and root password.

\noindent\ding{173}\ \textbf{System identity}: different hostname.

\noindent\ding{174}\ \textbf{Enumeration noise}: 3~non-exploitable SUID binaries, 2~decoy user accounts, and 2~harmless cron entries injected to pollute enumeration output, the most discriminative axis, forcing agents to distinguish genuinely exploitable from benign.

\noindent\ding{175}\ \textbf{Context noise}: a custom MOTD banner and a pre-seeded misleading shell history containing commands unrelated to the actual vulnerability (e.g., Docker, Kubernetes, database administration).

\noindent\ding{176}\ \textbf{Environment fingerprint}: additional common utilities for network and file inspection are pre-installed to alter the discoverable toolchain.

\begin{figure}[h]
    \centering
    \includegraphics[width=\columnwidth]{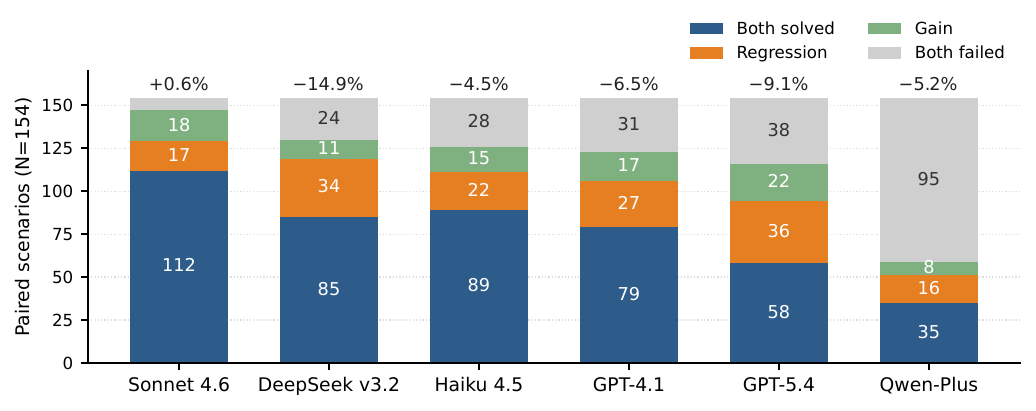}
    \caption{Original--variant outcome overlap on the shared 154-pair panel.}
    \Description{Stacked bars show both solved, regression, gain, and both failed outcomes for each model on the shared original--variant panel; labels above bars show the change in success rate.}
    \label{fig:vr-overlap}
\end{figure}

\noindent\textbf{Metric.} The primary metric is \textbf{retention}: among original scenarios a model solved in RQ1, the fraction it also solves after perturbation. We additionally report Pearson VR over the shared 154-pair panel and apply McNemar's test~\cite{mcnemar1947note} to the corresponding original--variant paired outcomes; \Cref{fig:vr-overlap} visualizes the shared-panel outcome overlap.

\noindent\textbf{Results.} \Cref{tab:results-vr} and \Cref{fig:vr-overlap} together reveal three behavioral patterns. \textbf{(1)~Retention is bounded even for strong models}: no model preserves all demonstrated successes; the best retention rates are Claude Haiku~4.5 (93/119, 78.2\%) and Claude Sonnet~4.6 (156/200, 78.0\%). \textbf{(2)~Reasoning capability alone does not ensure high retention}: GPT-5.4 leads RQ1 aggregate SR but has the lowest retention (128/217, 59.0\%), indicating dependence on environmental cues disrupted by the variants. \textbf{(3)~Retention and paired consistency capture different effects}: Qwen-Plus has high shared-panel VR (0.638) but low absolute capability, so we use retention as the primary perturbation metric.

\begin{table}[h]
\centering
\caption{Variant perturbation results per model.}
\label{tab:results-vr}
\small
\begin{tabular}{@{}lcccc@{}}
\toprule
\textbf{Model} & \textbf{RQ1 solved} & \textbf{Retained} & \textbf{Retention} & \textbf{VR} \\
\midrule
Claude Haiku~4.5 & 119 & 93 & \textbf{78.2\%} & 0.434 \\
Claude Sonnet~4.6 & 200 & 156 & 78.0\% & 0.151 \\
GPT-4.1 & 117 & 80 & 68.4\% & 0.374 \\
Qwen-Plus & 54 & 36 & 66.7\% & 0.638 \\
DeepSeek~v3.2 & 143 & 94 & 65.7\% & 0.346 \\
GPT-5.4 & 217 & 128 & 59.0\% & 0.244 \\
\bottomrule
\end{tabular}
\end{table}

\begin{tcolorbox}[colback=blue!5, colframe=blue!40!black]
\textbf{Finding 5:} Environmental perturbation removes 21.8--41.0\% of previously demonstrated model successes, so configuration rotation partially mitigates but does not eliminate automated escalation risk.
\end{tcolorbox}

\subsection{RQ3: Agent Architecture Impact}
\label{sec:rq3}

To test how agent architecture changes measured capability across models, we evaluate HackSynth~\cite{muzsai2024hacksynth} and wintermute~\cite{happe2026llms} over the full 531~scenarios on all six RQ1 models.

\begin{table}[h]
\centering
\caption{Architecture impact: wintermute vs HackSynth.}
\label{tab:rq3-arch}
\small
\begin{tabular}{@{}lccc@{}}
\toprule
\textbf{Model} & \textbf{wintermute SR} & \textbf{HackSynth SR} & $\Delta$ pp \\
\midrule
GPT-5.4 & 40.9\% & 47.3\% & \textbf{+6.4} \\
Claude Sonnet~4.6 & 37.7\% & \textbf{66.5\%} & \textbf{+28.8} \\
DeepSeek~v3.2 & 26.9\% & 19.2\% & $-7.7$ \\
GPT-4.1 & 22.0\% & 29.2\% & +7.2 \\
Claude Haiku~4.5 & 22.4\% & 33.0\% & +10.6 \\
Qwen-Plus & 10.2\% & 24.9\% & +14.7 \\
\bottomrule
\end{tabular}
\end{table}

\noindent\textbf{Results.} \Cref{tab:rq3-arch} reveals three interlocking patterns across the full six-model panel. \textbf{(1)~Planner-Summarizer uplift is model-dependent}: HackSynth improves five models, with the largest gains on Claude Sonnet~4.6 ($+28.8$\,pp), Qwen-Plus ($+14.7$\,pp), Claude Haiku~4.5 ($+10.6$\,pp), and GPT-4.1 ($+7.2$\,pp), but falls below wintermute for DeepSeek~v3.2 ($-7.7$\,pp). \textbf{(2)~Reasoning capability mutes the architectural uplift}: for GPT-5.4, the same architecture yields only $+6.4$\,pp; HackSynth-vs-wintermute gains are significant for Claude Sonnet~4.6 and Qwen-Plus (McNemar $p<0.001$) but not GPT-5.4 ($p=0.09$). \textbf{(3)~The aggregate ranking flips under HackSynth}: Claude Sonnet~4.6 (66.5\%) now leads GPT-5.4 (47.3\%) by 19.2~pp, the opposite of their wintermute ordering, so agent-architecture interventions must be evaluated jointly with model class.

\begin{tcolorbox}[colback=blue!5, colframe=blue!40!black]
\textbf{Finding 6:} Agent architecture materially changes automated escalation capability and can reorder aggregate model rankings; the size and direction of the uplift are model-dependent.
\end{tcolorbox}


\subsection{RQ4: Cost and Efficiency}
\label{sec:rq4-cost}

In addition to capability and robustness, the practical threat also depends on economic viability and the speed at which successful attacks conclude. \Cref{tab:results-cost} reports per-model cost and step efficiency from the available wintermute and HackSynth logs under a common list-price accounting method. A \textit{step} is one agent-environment interaction issuing a single shell command, so \textit{Steps/Succ.} is the mean number of agent-issued commands among successful runs, with lower values indicating faster root access. The wintermute ReAct pattern issues one LLM call per step that interleaves reasoning with the next action, while HackSynth's Planner-Summarizer architecture issues two LLM calls per step (a Planner proposes the next action and a Summarizer compresses the trace). Costs are computed using official per-token pricing for each model, with GPT-5.4's reasoning tokens additionally billed at the output rate.

\begin{table}[h]
\centering
\caption{Cost and step efficiency per agent and model.}
\label{tab:results-cost}
\resizebox{\columnwidth}{!}{%
\begin{tabular}{@{}llccccc@{}}
\toprule
\textbf{Agent} & \textbf{Model} & \textbf{Steps/Succ.} & \textbf{Avg In Tok.} & \textbf{Cost/Scen.} & \textbf{Total} & \textbf{Cost/Succ.} \\
\midrule
\multirow{6}{*}{wintermute}
& Claude Sonnet~4.6 & 9.0 & 52,506 & \$0.169 & \$88.14 & \$0.44 \\
& Claude Haiku~4.5 & \textbf{7.3} & 52,897 & \$0.044 & \$23.07 & \$0.20 \\
& GPT-4.1 & 9.7 & 94,362 & \$0.191 & \$99.91 & \$0.85 \\
& DeepSeek~v3.2 & 9.8 & 21,022 & \$0.006 & \$3.11 & \textbf{\$0.02} \\
& GPT-5.4 & 10.4 & 170,935 & \$0.500 & \$265.38 & \$1.22 \\
& Qwen-Plus & 9.6 & 67,120 & \$0.027 & \$14.30 & \$0.27 \\
\midrule
\multirow{6}{*}{HackSynth}
& Claude Sonnet~4.6 & \textbf{5.6} & 40,009 & \$0.237 & \$125.91 & \$0.36 \\
& Claude Haiku~4.5 & 7.2 & 58,329 & \$0.113 & \$59.90 & \$0.34 \\
& GPT-4.1 & 6.5 & 42,419 & \$0.144 & \$76.57 & \$0.49 \\
& DeepSeek~v3.2 & 7.3 & 38,179 & \$0.032 & \$17.04 & \$0.17 \\
& GPT-5.4 & 6.9 & 38,112 & \$0.722 & \$383.36 & \$1.53 \\
& Qwen-Plus & 10.7 & 47,669 & \$0.029 & \$15.28 & \textbf{\$0.11} \\
\bottomrule
\end{tabular}}
\end{table}

\noindent\textbf{Cost.} Under the wintermute baseline, per-success cost spans \$0.02 (DeepSeek~v3.2 at 26.9\% SR) to \$1.22 (GPT-5.4 at 40.9\% SR), indicating that API cost is low relative to the cost of running full interactive evaluations. The top non-reasoning model, Claude Sonnet~4.6 at 37.7\% SR, costs \$0.44 per success; reasoning-augmented GPT-5.4 costs nearly $3\times$ that. Lower-cost models (DeepSeek~v3.2, Qwen-Plus) achieve 27 to 71\% of Claude Sonnet~4.6's SR at 5 to 60\% of its per-success cost. HackSynth shifts the per-success picture: cost drops for Claude Sonnet~4.6 (\$0.44 to \$0.36) and Qwen-Plus (\$0.27 to \$0.11) thanks to higher SR, but rises for GPT-5.4 to \$1.53 because reasoning-token billing scales with the larger output volume produced per step.

\noindent\textbf{Step-budget efficiency.} Successful attacks terminate well before the budget is exhausted: Claude Haiku~4.5 and Claude Sonnet~4.6 end in 7.3 and 9.0 commands on average, with \textbf{29.4\% of Claude Haiku~4.5 and 23.5\% of Claude Sonnet~4.6 successes completing in $\leq 3$ commands}, leaving fewer interaction steps for runtime detection. GPT-5.4, despite the highest overall SR, is the slowest per success (10.4 commands, only 4.1\% within 3 commands and 21.2\% using 16 to 20 commands), consistent with reasoning models exploring longer chains before committing. By contrast, failed runs that exhaust the budget cost roughly $4\times$ as much per scenario as successful ones, explaining the gap between Cost/Scen and Cost/Succ in \Cref{tab:results-cost}. Within wintermute, the cost-vs-speed frontier has two extremes: Claude Haiku~4.5 is fastest and second-cheapest per success (7.3 commands, \$0.20); DeepSeek~v3.2 is cheapest but slightly slower (9.8 commands, \$0.02).

\begin{tcolorbox}[colback=blue!5, colframe=blue!40!black]
\textbf{Finding 7:} Successful runs are often short and have low measured token cost under our setup, reducing the time available for runtime detection.
\end{tcolorbox}

\begin{tcolorbox}[colback=blue!5, colframe=blue!40!black]
\textbf{Finding 8:} Per-success token cost depends strongly on the agent architecture, not just the underlying model.
\end{tcolorbox}

\section{PrivEscAgent: Domain-Specialized Threat Amplification}
\label{sec:privescagent}

Our measurements show that agent architecture can substantially change measured success rates, with the effect varying by model. This raises a follow-up question: \emph{how much can a domain-specialized agent change measured privilege-escalation capability?} Motivated by this question and the bottlenecks observed in RQ1 and RQ2, we design \textsc{PrivEscAgent} as a domain-specialized measurement probe: a wrapper that addresses each bottleneck with a targeted module.

\subsection{Key Bottlenecks}
\label{sec:reflection}

Our measurement reveals key bottlenecks in current agent-based privilege escalation. \textbf{Blind enumeration}: agents waste most of their step budget on unfocused exploration before identifying the vulnerability class. \textbf{Missing domain knowledge}: agents discover the vulnerability class but lack exploit-specific knowledge, attempting wrong techniques even on straightforward misconfigurations. \textbf{Multi-step breakdown}: agents often struggle on exploits requiring multiple coordinated steps (e.g., cron job timing, file write with privilege switch). \textbf{Strategy fixation}: agents persist on the initially chosen strategy, cycling its variants rather than pivoting to an alternative exploitation category. These bottlenecks motivate the design of PrivEscAgent, a lightweight wrapper around the ReAct base agent that addresses each with a targeted module.

\subsection{Architecture}

\begin{figure}[t]
    \centering
    \includegraphics[width=\linewidth]{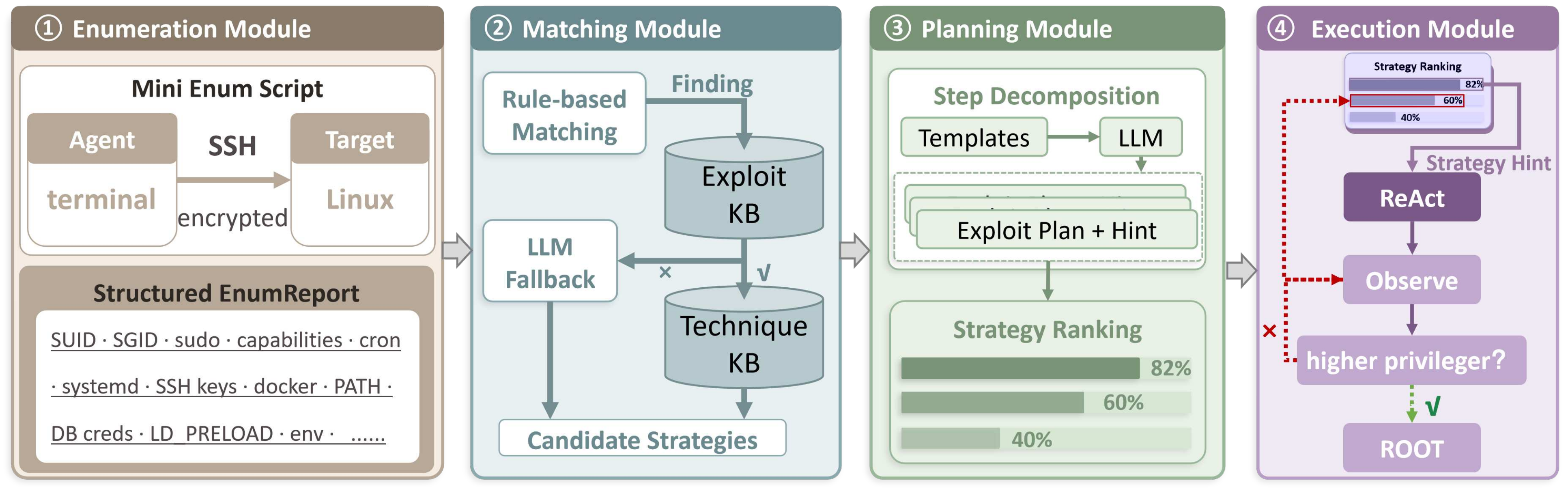}
    \caption{\textsc{PrivEscAgent} architecture.}
    \Description{Architecture diagram showing PrivEnum, CategoryMatcher, StrategySelector, StepPlanner, and the ReAct execution loop.}
    \label{fig:privescagent}
\end{figure}

As \Cref{fig:privescagent} shows, PrivEscAgent extends wintermute with four modules: PrivEnum, CategoryMatcher, StrategySelector, and StepPlanner. PrivEnum performs automated enumeration; CategoryMatcher maps its output to ATT\&CK categories; StrategySelector ranks candidate strategies and auto-pivots on execution failure; StepPlanner decomposes the chosen strategy into verifiable steps. These modules run as a preprocessing pipeline, producing an exploitation plan that is executed by wintermute's ReAct loop.

\subsubsection{PrivEnum.} PrivEnum is a deterministic, LLM-free enumeration module that addresses the \textbf{blind enumeration} bottleneck. Once the agent obtains an initial low-privilege foothold, PrivEnum executes a single compact script through that shell in one round-trip, collecting taxonomy-aligned privilege-escalation evidence: SUID/SGID binaries, sudo rules, Linux capabilities, cron entries, and the remaining vectors in \Cref{tab:taxonomy}. This consolidates what would otherwise span many unfocused enumeration steps into a single deterministic command, freeing the remaining step budget for exploitation.

\subsubsection{CategoryMatcher.} CategoryMatcher is a two-tier classification module that addresses the \textbf{missing domain knowledge} bottleneck. It maps PrivEnum's output to the most likely ATT\&CK categories using a rule-based knowledge-base lookup for documented exact matches, falling back to an LLM-assisted reasoner for ambiguous or uncovered cases. The module emits a ranked list of candidate exploitation strategies.

\subsubsection{StepPlanner.} StepPlanner is a decomposition module that addresses the \textbf{multi-step breakdown} bottleneck, activated only when exploitation requires chained steps. Given the identified category and candidate strategy, it invokes an LLM to decompose the exploit into verifiable steps with rollback to the previous step on failure: identify the exploitable target, inject the payload, await the trigger, and verify root access. For single-step exploits, this module is skipped.

\subsubsection{StrategySelector.} StrategySelector is a ranking-and-pivot module that addresses the \textbf{strategy fixation} bottleneck. When CategoryMatcher produces multiple candidate strategies, it ranks them using CategoryMatcher's confidence weighted by per-category success-rate priors, and on execution failure automatically pivots to the next candidate without restarting the episode, reducing wasted steps on incorrect strategies.

\subsection{Effectiveness}
\label{sec:privescagent-eval}

\noindent\textbf{Setup.} We evaluate \textsc{PrivEscAgent} on \textsc{PrivEscalate} against the two RQ3 baselines (wintermute, HackSynth) using all six models from the RQ1/RQ2 panel. CategoryMatcher's knowledge base is instantiated with GTFOBins~\cite{gtfobins}; StrategySelector combines its per-candidate confidence with per-category SR priors from RQ1 and RQ2.

\begin{table}[h]
\centering
\caption{Baseline comparison across agents.}
\label{tab:baseline}
\small
\begin{tabular}{@{}lccc@{}}
\toprule
\textbf{Model} & \textbf{wintermute} & \textbf{HackSynth} & \textbf{PrivEscAgent} \\
\midrule
GPT-5.4 & 40.9 & 47.3 & \textbf{54.6} \\
Claude Sonnet~4.6 & 37.7 & 66.5 & \textbf{71.9} \\
DeepSeek~v3.2 & 26.9 & 19.2 & \textbf{42.9} \\
GPT-4.1 & 22.0 & 29.2 & \textbf{47.8} \\
Claude Haiku~4.5 & 22.4 & 33.0 & \textbf{45.6} \\
Qwen-Plus & 10.2 & 24.9 & \textbf{39.5} \\
\bottomrule
\end{tabular}
\end{table}

\noindent\textbf{Baseline comparison.} \Cref{tab:baseline} shows that across all six models, PrivEscAgent outperforms both prior frameworks. The most pronounced relative change is Qwen-Plus, the lowest-SR model under wintermute (10.2\%): with PrivEscAgent it reaches 39.5\% SR, comparable to Claude Sonnet~4.6's 37.7\% wintermute baseline. GPT-4.1, Claude Haiku~4.5, and DeepSeek~v3.2 also rise substantially under PrivEscAgent, showing that agent design can materially affect measured capability.

\noindent\textbf{Per-model results.} Across the six-model panel, PrivEscAgent is the top-SR framework for every model, while the ordering between the two prior baselines is model-dependent and even reverses for DeepSeek. The largest absolute gain over wintermute occurs for Claude Sonnet~4.6 ($+34.3$\,pp), followed by Qwen-Plus ($+29.4$\,pp), GPT-4.1 ($+25.8$\,pp), Claude Haiku~4.5 ($+23.2$\,pp), DeepSeek~v3.2 ($+16.0$\,pp), and GPT-5.4 ($+13.7$\,pp). Category-level inspection shows that the gains concentrate in common privilege-escalation families such as sudo, capabilities, and credentials, while SUID/SGID improvements are more model-dependent.

\subsection{Cost and Strategy Compliance}
\label{sec:privescagent-cost}

\noindent\textbf{Cost analysis.} \Cref{tab:privescagent-cost} reports per-success cost for PrivEscAgent and its relative change against the baseline cost rows in \Cref{tab:results-cost}. PrivEscAgent lowers per-success cost relative to HackSynth for all six models and relative to wintermute for five models; the exception is DeepSeek~v3.2, whose wintermute run is already very low-cost.

\begin{table}[h]
\centering
\caption{Per-success cost: PrivEscAgent vs.\ baselines.}
\label{tab:privescagent-cost}
\small
\resizebox{\columnwidth}{!}{%
\begin{tabular}{@{}lccc@{}}
\toprule
\textbf{Model} & \textbf{PrivEscAgent} & \textbf{$\Delta$ vs.\ wintermute} & \textbf{$\Delta$ vs.\ HackSynth} \\
\midrule
Claude Sonnet~4.6 & \$0.17 & $-61\%$ & $-53\%$ \\
Claude Haiku~4.5 & \$0.12 & $-40\%$ & $-65\%$ \\
GPT-4.1 & \$0.17 & $-80\%$ & $-65\%$ \\
DeepSeek~v3.2 & \$0.06 & $+200\%$ & $-65\%$ \\
Qwen-Plus  & \$0.03 & $-89\%$ & $-73\%$ \\
GPT-5.4    & \$0.14 & $-89\%$ & $-91\%$ \\
\bottomrule
\end{tabular}}
\end{table}

\noindent\textbf{Strategy compliance.} We also examine how models use the same GTFOBins-derived strategy hints. The behavior is model-dependent: Qwen-Plus succeeds on \textbf{40 scenarios where GPT-5.4 fails} and \textbf{8 where Claude Sonnet~4.6 fails} despite receiving identical hints. Among these failures, 32\% for GPT-5.4 and 75\% for Claude Sonnet~4.6 exhaust the 20-step budget, suggesting that the model's chosen execution path did not converge within the allowed budget. On co-success scenarios, Claude Sonnet~4.6 uses $\geq 3$ extra steps on 27.9\% of runs (mean $+1.6$ vs.\ Qwen-Plus), while GPT-5.4 does so on 18.0\% (mean $+0.1$).

\noindent\textbf{Case studies.} On \texttt{sudo\_check\_by\_ssh}, Claude Sonnet~4.6 substitutes a shell it deems more portable and removes a TTY (terminal-control) redirection that sudo requires, exhausting 20 steps, while Qwen-Plus follows the hint and reaches root in 4. On \texttt{sudo\_at}, GPT-5.4 modifies the hint and exhausts 20 steps while Qwen-Plus succeeds in 4. These examples illustrate a design trade-off for agent-knowledge-base integration: stronger adherence can preserve verified strategies, but overly rigid execution may limit useful adaptation outside the knowledge base.

\begin{tcolorbox}[colback=blue!5, colframe=blue!40!black]
\textbf{Finding 9:} Strategy-hint use is model-dependent; verified hints improve the agent pipeline, but models differ in how closely they follow them.
\end{tcolorbox}

\subsection{Ablation Study}
\label{sec:privescagent-ablation}

\noindent\textbf{Setup.} We evaluate four configurations on all six RQ1 models, listed in \Cref{tab:ablation}: the full PrivEscAgent (PrivEnum + CategoryMatcher / CM + StepPlanner / SP), and three ablations that remove SP, both CM and SP, or PrivEnum. StrategySelector activates only when CategoryMatcher emits multiple candidates and is folded into the CM condition rather than ablated separately.

\begin{figure}[t]
    \centering
    \begin{tikzpicture}
    \begin{axis}[
        width=\linewidth,
        height=4.6cm,
        ybar,
        ymin=-22,
        ymax=6,
        ylabel={$\Delta$SR vs. full (pp)},
        symbolic x coords={Sonnet,Qwen+,GPT-5.4,GPT-4.1,Haiku,DeepSeek},
        xtick=data,
        x tick label style={rotate=35, anchor=east, font=\scriptsize},
        ymajorgrids=true,
        grid style={dotted, gray!35},
        bar width=4pt,
        enlarge x limits=0.12,
        legend style={font=\scriptsize, at={(0.5,1.04)}, anchor=south, legend columns=3, draw=none}
    ]
    \addplot+[fill=blue!55!black, draw=blue!55!black] coordinates {(Sonnet,-7.5) (Qwen+,-11.4) (GPT-5.4,4.5) (GPT-4.1,1.9) (Haiku,2.3) (DeepSeek,0.6)};
    \addplot+[fill=orange!80!black, draw=orange!80!black] coordinates {(Sonnet,-8.4) (Qwen+,-15.0) (GPT-5.4,2.7) (GPT-4.1,0.8) (Haiku,-0.8) (DeepSeek,-1.3)};
    \addplot+[fill=red!65!black, draw=red!65!black] coordinates {(Sonnet,-19.2) (Qwen+,-19.0) (GPT-5.4,0.0) (GPT-4.1,-8.7) (Haiku,-5.3) (DeepSeek,-5.8)};
    \legend{$-$SP,$-$CM\&SP,$-$PrivEnum}
    \end{axis}
    \end{tikzpicture}
    \caption{Ablation impact across all six evaluated models. Values are percentage-point changes in SR relative to full PrivEscAgent.}
    \Description{Grouped bar chart showing success-rate changes for removing StepPlanner, CategoryMatcher with StepPlanner, or PrivEnum across six models.}
    \label{fig:ablation-overlap}
\end{figure}
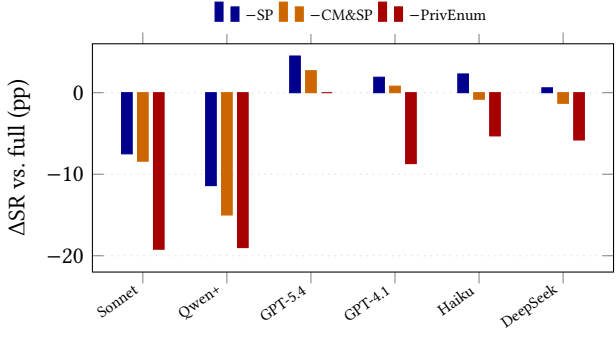

\noindent\textbf{Results.} \Cref{tab:ablation} and \Cref{fig:ablation-overlap} show three module-level patterns across the six-model panel. First, PrivEnum is the most consistent contributor: removing it causes the largest SR drop for five non-reasoning models, ranging from $-5.3$ to $-19.2$\,pp. Second, StepPlanner and CategoryMatcher are useful but not universally positive: removing StepPlanner hurts Claude Sonnet~4.6 and Qwen-Plus substantially, has little effect on GPT-4.1, Claude Haiku~4.5, and DeepSeek~v3.2, and improves GPT-5.4. Third, the $-$CM\&SP condition measures performance without CategoryMatcher, StrategySelector, or category-level SR priors; Claude Sonnet~4.6 still reaches 63.5\% SR, $+25.8$\,pp above wintermute.

\noindent\textbf{Module effects.} PrivEnum is the dominant component for the non-reasoning models: removing it drops Claude Sonnet~4.6 and Qwen-Plus by about 19\,pp and also produces the largest drop for GPT-4.1, Claude Haiku~4.5, and DeepSeek~v3.2. StepPlanner contributes most clearly for Claude Sonnet~4.6 ($-7.5$\,pp) and Qwen-Plus ($-11.4$\,pp), indicating that explicit step decomposition helps models that otherwise struggle to convert a matched strategy into executable commands. CategoryMatcher and StrategySelector add smaller aggregate gains after StepPlanner is removed; their ablation also reports performance without category-level SR priors.

\noindent\textbf{Model-level variation.} GPT-5.4 behaves differently from the non-reasoning models: removing StepPlanner or both CM and SP increases aggregate SR, while removing PrivEnum leaves aggregate SR unchanged. The overlap analysis in \Cref{fig:ablation-overlap} shows that these aggregate changes can hide exchanges between different success sets. Thus, the same scaffolding can amplify weaker models while reshaping, rather than strictly increasing, the behavior of a reasoning-augmented model.

\begin{table}[h]
\centering
\caption{\textsc{PrivEscAgent} ablation across all six RQ1 models (SR \%).}
\label{tab:ablation}
\scriptsize
\setlength{\tabcolsep}{3.1pt}
\begin{tabular}{@{}lrrrrrr@{}}
\toprule
\textbf{Config.} & \textbf{Sonnet} & \textbf{Qwen+} & \textbf{GPT-5.4} & \textbf{GPT-4.1} & \textbf{Haiku} & \textbf{DeepSeek} \\
\midrule
Full    & \textbf{71.9} & \textbf{39.5} & 54.6 & 47.83 & 45.57 & 42.94 \\
$-$SP   & 64.4 & 28.1 & \textbf{59.1} & \textbf{49.72} & \textbf{47.83} & \textbf{43.50} \\
$-$CM\&SP      & 63.5 & 24.5 & 57.3 & 48.59 & 44.82 & 41.62 \\
$-$PrivEnum      & 52.7 & 20.5 & 54.6 & 39.17 & 40.30 & 37.10 \\
\bottomrule
\end{tabular}
\vspace{1mm}
\end{table}

\noindent\textbf{Statistical interpretation.} McNemar tests identify PrivEnum as the most consistent contributor, with significant drops for GPT-4.1 and DeepSeek~v3.2 and a similar direction for Claude Haiku~4.5. StepPlanner and CM\&SP effects vary more across models, indicating that decomposition and category-guided ranking act as conditional scaffolds whose benefit depends on the base model.

\section{Discussion}
\label{sec:discussion}

\subsection{Adversarial Capability Limits}
\label{sec:capability-limits}

We manually analyze the 61 of 531 scenarios (11.5\%) that remain unsolved across all evaluated model-agent configurations; the inspected cases point to capability gaps rather than environment defects. This set is distinct from the 202 scenarios unsolved by any of the six models under wintermute alone: adding HackSynth and PrivEscAgent solves 141 of those wintermute-unsolved scenarios. Two failure patterns dominate the final unsolved set: (1) 31 scenarios (51\%) require a controlling TTY or interactive prompt that agents fail to allocate, such as those needed by \texttt{aspell} or \texttt{ed}; and (2) 24 scenarios (39\%) require multi-step credential workflows such as reading \texttt{/etc/shadow}, cracking a weak password offline, and then invoking \texttt{su}, which exceed the agents' multi-step planning and tool-chaining capabilities. SUID/SGID accounts for 61\% of the final unsolved set. The observed TTY and credential patterns correspond to two standard hardening measures for the affected deployment classes: require interactive TTYs for commands that need terminal control, and use password hashes resistant to offline cracking.

\subsection{Defensive Recommendations}
\label{sec:defense}

We organize defensive recommendations into four layers grounded in our measurements.

\noindent\textbf{Configuration hardening.} At the configuration layer, three actions are aligned with the highest-risk categories measured in our benchmark: eliminating sudoers entries that grant password-free root execution, restricting SUID/SGID binaries to a minimal vetted set, and removing sensitive Linux capabilities (such as those that allow user-ID switching or privileged file reads) from non-root binaries. These categories show the highest measured exploitation success rates across all six evaluated models in our benchmark and therefore merit priority. Because some models retain exploitation success under environmental perturbation, configuration rotation alone is incomplete and should be paired with stronger access-control policies.

\noindent\textbf{Behavioral detection.} Three SOC-consumable detection signatures emerge from our failed-run traces, where enumeration commands dominate across all evaluated models: alerts on bursts of enumeration calls within tight time windows, on repeated invocations of the same binary with mutated flag combinations, and on sessions whose command stream is anomalously uniform relative to human administrator baselines.

\noindent\textbf{Architectural controls.} Mandatory access control (MAC) is an important complement to configuration hardening because it enforces system-level policies that bound what any process can do independent of the exploitation path. This property is useful when environmental variation or agent design changes the exact command sequence used during escalation. Detection rules should also be evaluated against augmented agents, since rules tuned only to baseline ReAct behavior may underestimate activity patterns produced by domain-specialized agents.

\noindent\textbf{Operational scanning and audit.} Recurring scans of SUID/SGID binaries, sudoers entries, Linux capabilities, and writable cron and systemd unit files surface configuration drift before adversaries can exploit it; CVE-based scanning of installed packages catches privilege escalation primitives tied to known software vulnerabilities. Asset inventory should track every privileged binary, service unit, and authentication-relevant configuration file across the deployment, with comparisons against a known-good baseline triggering alerts on unexpected additions or permission changes. Periodic audit-log review further detects manual configuration changes that bypass infrastructure-as-code controls.

\subsection{Limitations}
\label{sec:limitations}

\noindent\textbf{Taxonomy coverage.} The prior hackingBuddyGPT benchmark excluded kernel exploits (host instability under repeated runs), NFS root squashing (which requires a dedicated remote attacker host), and service-specific exploits (because of product and version dependence), and it omitted weak file-system permissions on sensitive system files. PrivEscalate retains the first three exclusions on those grounds but re-includes weak file-system permissions through two expert-authored scenarios because they remain operationally relevant in misconfigured deployments encountered in audits. The result is a 14-category Linux privilege-escalation taxonomy, with per-category rationale in \Cref{app:excluded-categories}. These 14 sub-categories cover widely documented Linux privilege-escalation vectors represented in GTFOBins, Exploit-DB, and penetration-testing curricula.

\noindent\textbf{Statistical scope.} Four factors shape interpretation of the aggregate measurements. \textbf{(1)~Variant correlation}: variant scenarios within the same template are inherently correlated despite the original-variant pairing design, so SR on the original corpus and retention under perturbation are analyzed separately; Pearson VR in \Cref{tab:results-vr} reports paired consistency. \textbf{(2)~Source-driven category distribution}: SUID/SGID (38\%) and Sudo (55\%) dominate the 14 sub-categories because GTFOBins and Exploit-DB contain more reproducible entries for these classes; per-sub-category reporting exposes how this common-case distribution relates to class-specific behavior. \textbf{(3)~Run-to-run variation}: most model-agent-scenario tuples are evaluated once. A separate one-run wintermute repeatability check preserved the primary qualitative model ordering. Paired McNemar tests support the original--variant, architectural, and ablation comparisons. \textbf{(4)~Evaluation scope}: the 20-step cap follows prior automated Linux privilege-escalation agent studies and supports comparability, but may over-allocate easy scenarios and understate longer-chain cases. The study is designed for relative LLM-agent comparison rather than human-agent comparison.

\noindent\textbf{Agent framework coverage and tuning.} We evaluate three agents spanning the major paradigms in current LLM-based penetration testing: wintermute as the ReAct baseline, HackSynth as the Planner-Summarizer baseline, and PrivEscAgent as our domain-specialized augmentation of ReAct. The fixed-budget protocol requires transparent step control, so frameworks with opaque or non-configurable control loops are left to future benchmark extensions. StrategySelector uses category-level SR priors computed from earlier measurements; the $-$CM\&SP ablation removes CategoryMatcher, StrategySelector, and those priors to measure performance without category-prior ranking. The benchmark's SSH-based interface is framework-agnostic and supports any agent capable of executing shell commands, facilitating future evaluation of additional architectures.

\section{Related Work}
\label{sec:related}

\subsection{LLM Agents for Offensive Security}

Recent advances in LLMs have enabled agents that combine reasoning and environment interaction, as exemplified by ReAct~\cite{yao2023react}. Inspired by such paradigms, prior work has explored LLM-based agents for offensive security tasks. Existing systems can be broadly categorized into two groups, distinguished primarily by task granularity and the degree of human oversight required during execution. \textbf{Workflow-oriented agents} orchestrate multi-stage penetration testing through planning and tool use, as exemplified by PentestGPT~\cite{deng2024pentestgpt}, which employs a multi-module design that decomposes the attack process into separate reasoning, command generation, and output parsing stages. \textbf{Capability-oriented agents} focus on the autonomous execution of specific attack primitives. Perses~\cite{cichon2025perses} proposes a multi-LLM framework for misconfiguration-based privilege escalation and evaluates it on FreeBSD systems, while Fang et al.~\cite{fang2024llm_2} show that LLM agents can autonomously discover and exploit web vulnerabilities without human intervention. Predating these LLM-based approaches, ChainReactor~\cite{depasquale2024chainreactor} uses classical PDDL planning to discover privilege escalation chains on real systems, but requires manual encoding of attack actions and CVE-specific predicates, which limits its scalability across diverse Linux distributions and exploit families.
Despite these advances, current approaches are typically evaluated in restricted or task-specific environments, with limited reproducibility, fragmented coverage of offensive tasks, and inconsistent evaluation protocols across systems. This makes it difficult to systematically compare model capabilities or assess robustness across scenarios, motivating the need for reproducible, scalable, and verifiable evaluation frameworks for system-level tasks such as Linux privilege escalation.

\subsection{Security Benchmarks for LLM Evaluation}

Alongside the development of LLM-based offensive agents, recent work has proposed benchmarks to evaluate their capabilities. Early efforts largely rely on CTF-style tasks drawn from public security competitions, such as Cybench and NYU CTF Bench, which provide scalable evaluation settings with executable environments~\cite{zhang2025cybench, shao2024nyuctf}. To improve realism, subsequent benchmarks introduce more structured and practical settings: AutoPenBench~\cite{gioacchini2025autopenbench} provides executable multi-stage penetration testing tasks with milestone-based evaluation, Isozaki et al.~\cite{isozaki2025towards} construct an end-to-end VM-based benchmark with detailed analysis of LLM limitations, and PentestEval~\cite{yang2025pentesteval} further introduces a modular, stage-level benchmark design for fine-grained evaluation. In parallel, CVE-Bench and SEC-bench focus on real-world vulnerabilities, providing executable environments with exploit validation and reproducible evaluation protocols~\cite{wen2025cvebench, maier2025secbench}.

Despite this progress, no existing benchmark provides systematic coverage of Linux privilege escalation at scale. HackingBuddyGPT~\cite{happe2026llms} is the only dedicated effort, but its 13~scenarios are insufficient to support statistically meaningful cross-model comparisons or perturbation-sensitivity analysis. PrivEscalate addresses this gap with 531~audited, Dockerized scenarios spanning 14 ATT\&CK-mapped sub-categories.

\subsection{Exploit Environment Construction}

Beyond benchmark design, recent work explores approaches to constructing reproducible security evaluation environments. Agent-oriented systems such as hackingBuddyGPT~\cite{happe2026llms} evaluate Linux privilege escalation in virtual machines, whereas Perses~\cite{cichon2025perses} evaluates misconfiguration-based privilege escalation on FreeBSD systems. SEC-bench and CVE-Bench construct Dockerized environments from real-world vulnerabilities with executable exploit verification~\cite{maier2025secbench, wen2025cvebench}; DrillAgent and VulnSage incorporate runtime feedback mechanisms to refine exploit generation based on execution behavior~\cite{li2026execution, chen2026multi}.
However, existing pipelines either rely on task-specific or partially manual environment construction, or focus on web vulnerabilities and general CVE exploitation. None address the challenges of large-scale Linux privilege escalation construction, including automated dependency installation when materializing Linux vulnerability scenarios, differential verification across vulnerable and patched containers, and systematic coverage across ATT\&CK sub-categories within a single automated pipeline. PrivEscalate and PrivEscAgent are designed to fill this gap.

\section{Conclusion}
\label{sec:conclusion}

We present \textsc{PrivEscalate}, a large-scale Linux privilege escalation testbed for LLM agent evaluation, defensive tool validation, and red-team training. Our measurements reveal that automated escalation capability is shaped jointly by model, environment, and agent architecture, rather than by raw model capability alone. Different models are strongest on different vulnerability classes, most models that appear capable on clean scenarios degrade once surface details of the environment change, and the agent framework can be as consequential as the model itself. Motivated by these findings, we develop \textsc{PrivEscAgent}, a domain-specialized wrapper that raises exploitation success for all six evaluated models without altering the underlying LLM. These results suggest that defenders should evaluate both model and agent design, since agent architecture can change measured capability and model rankings. We release \textsc{PrivEscalate} as an open, Dockerized testbed so that this measurement can continue as models, agents, and defenses advance in concert.

\begin{acks}
This research is supported by the Nanyang Technological University Centre for Computational Technologies in Finance (NTU-CCTF) and the RIE2025 Industry Alignment Fund – Industry Collaboration Projects (IAF-ICP) (Award I2301E0026), administered by A*STAR, as well as supported by Alibaba Group and NTU Singapore through Alibaba-NTU Global e-Sustainability CorpLab (ANGEL).
Any opinions, findings, and conclusions or recommendations expressed
in this material are those of the author(s) and do not necessarily reflect the views of NTU-CCTF and ANGEL.
\end{acks}

\bibliographystyle{ACM-Reference-Format}
\balance
\bibliography{references}

\newpage
\appendix

\section{Open Science}
\label{app:open-science}
\label{app:camera-ready-notes}

The artifact release at \url{https://github.com/yxsec/PrivEscalate} contains the 531 audited Docker-based scenarios, 329 variants, the multi-agent construction pipeline, evaluation adapters, manifests, prompts, and environment specifications.

\section{Ethical Considerations}
\label{app:ethics}

\noindent\textbf{Existing knowledge.} All scenarios use well-known privilege escalation techniques publicly documented in GTFOBins~\cite{gtfobins}, Exploit-DB~\cite{exploitdb}, and standard penetration testing curricula (e.g., OSCP, HackTheBox); no novel vulnerabilities or zero-day exploits are disclosed, so the benchmark consolidates existing public knowledge into a structured evaluation format without lowering the barrier to attack.

\noindent\textbf{Containment.} All experiments execute exclusively in isolated Docker containers with no external network access; containers are ephemeral with minimal base images and no sensitive data, the SSH interface is bound to \texttt{localhost} only, and we verified that no container escape is possible through the evaluated vulnerability classes under our Docker configuration.

\noindent\textbf{Dual-use risk and responsible release.} Publishing exploitation success rates and agent architectures could inform adversarial use; however, following the established precedent of SEC-bench~\cite{maier2025secbench}, CVE-Bench~\cite{wen2025cvebench}, and Cybench~\cite{zhang2025cybench}, transparent measurement of automated attack capabilities supports informed defense, and our defensive discussion (\Cref{sec:defense}) maps the measured risks to corresponding hardening directions. We ask users to follow responsible disclosure practices and use the benchmark solely for defensive research, education, and authorized security testing.

\section{Datasheet for PrivEscalate}
\label{app:datasheet}

\noindent\textbf{Composition.} Each instance is a Docker-based Linux system with one injected privilege escalation vulnerability, a low-privilege user account, and a ground-truth exploit script. The dataset contains 860 instances: 531 audited original scenarios (523 audit-retained, 2 weak-permissions, 6 expert-authored) plus 329 variants, spanning all 14 sub-categories of the taxonomy.

\noindent\textbf{Collection Process.} Construction uses our multi-agent pipeline (\Cref{sec:privescgen}): DataIngester for vulnerability discovery, reference-guided LLM Dockerfile generation, and triple verification (build, exploit differential with \texttt{uid=0}, consistency check). Seed templates were designed by one security researcher; the rest is automated, with 100\% expert review coverage.

\noindent\textbf{Uses.} The dataset supports LLM agent architecture comparison, privilege escalation defense evaluation, security education, and reinforcement-learning training.

\noindent\textbf{Distribution and Maintenance.} Released open-source under a permissive license upon publication acceptance, with long-term archival preservation. The authors maintain the dataset and accept community contributions; the template-based architecture enables extensions following the established format and verification pipeline.

\section{Excluded Sub-techniques and Rationale}
\label{app:excluded-categories}

\Cref{tab:excluded-categories} lists the five groups of ATT\&CK techniques considered but excluded from \textsc{PrivEscalate}'s 14-subcategory taxonomy, with rationale for each. Because ATT\&CK techniques are coarse-grained, excluded subsets such as kernel or vendor-specific T1068 cases are distinct from retained user-space misconfiguration rows that share the same ATT\&CK identifier.

\begin{table}[h]
\centering
\caption{Excluded sub-techniques.}
\label{tab:excluded-categories}
\scriptsize
\setlength{\tabcolsep}{3pt}
\begin{tabular}{@{}p{1.5cm}p{1.7cm}p{3.2cm}@{}}
\toprule
\textbf{Group} & \textbf{ATT\&CK mapping / scope} & \textbf{Rationale} \\
\midrule
Kernel exploits      & T1068 subset            & Shares the host kernel; needs VM-level isolation and is unstable in benchmarks~\cite{happe2026llms}. \\
\addlinespace
NFS root squashing   & NFS-specific scope      & Requires a dedicated attacker host beyond single-container scope; no dedicated ATT\&CK sub-technique is used here. \\
\addlinespace
Service-specific CVE exploits & T1068 (service/software CVEs) & Require a particular daemon, appliance, or software version and are not portable across our single-container benchmark. \\
\addlinespace
Process injection    & T1055.008 / .009 / .014 & Requires specific runtime state on the target process; not deterministic. \\
\addlinespace
Boot/logon triggered & T1547, T1037            & Depends on boot or login lifecycle state that is not deterministic in an ephemeral container; scheduled-task and systemd-timer cases are retained when their trigger is explicit and testable. \\
\bottomrule
\end{tabular}
\end{table}

\section{Per-Category Exploitation Patterns}
\label{app:categories}

Per-sub-category success rates are reported in \Cref{tab:per-sub-cat-sr}; here we summarize the corresponding trigger conditions and exploitation ideas observed in the benchmark. Ground-truth exploitation scripts for every scenario ship with the artifact.

\noindent\textbf{SUID/SGID abuse (T1548.001).} The trigger is a binary with the setuid-root bit set whose intended functionality also permits a shell escape or arbitrary file read. Exploitation invokes the binary with arguments or sub-commands that trigger a shell escape or privileged file operation while retaining its elevated execution context.

\noindent\textbf{Sudo misconfiguration (T1548.003).} The trigger is a sudoers entry that grants a user the right to run one or more commands as root with no password, or that uses permissive wildcards. Exploitation first inspects the sudoers listing, then picks a binary from that listing whose arguments allow a shell escape or privileged file modification.

\noindent\textbf{Capabilities abuse (T1068).} The trigger is a fine-grained Linux file capability such as \texttt{cap\_setuid} or \texttt{cap\_dac\_read\_search} attached to a non-root binary. Exploitation invokes the binary in a way that exercises the capability, for example by loading a language runtime that explicitly changes its effective user ID or reads privileged files on behalf of the caller.

\noindent\textbf{Polkit misconfiguration (T1548).} The trigger is an overly permissive polkit rule that grants a low-privilege user unconditional authorization to invoke a privileged action. Exploitation issues the authorized action through \texttt{pkexec} or the equivalent polkit client, obtaining root execution without any additional authentication step.

\noindent\textbf{D-Bus misconfiguration (T1068).} The trigger is a system D-Bus service that exposes a privileged method without verifying the caller's identity. Exploitation issues a D-Bus call to the unprotected method, causing the service (which itself runs as root) to perform the privileged action on the attacker's behalf.

\noindent\textbf{Weak file permissions (T1222.002).} The trigger is an operator mistake leaving \texttt{/etc/passwd} or \texttt{/etc/shadow} world-writable. Exploitation clears root's password field (or appends a new UID-0 entry) and then authenticates with an empty password, relying on PAM's \texttt{nullok} option on the stock distribution.

\noindent\textbf{PATH hijacking (T1574.007).} The trigger is a user-writable directory that appears earlier in \texttt{\$PATH} than the system directories, often because a sudoers rule explicitly preserves PATH across \texttt{sudo}. Exploitation drops a malicious executable with the same name as a system utility into the writable directory, so the next privileged invocation resolves through the attacker's file.

\noindent\textbf{LD\_PRELOAD hijack (T1574.006).} The trigger is a sudoers or service configuration that preserves \texttt{LD\_PRELOAD} across privilege boundaries. Exploitation loads an attacker-supplied shared library into a privileged process; the library's constructor runs as root and spawns a shell.

\noindent\textbf{Cron job exploitation (T1053.003).} The trigger is a root-owned cron entry whose script (or the directory containing it) is writable by a lower-privilege user. Exploitation modifies the script to perform an attacker-chosen action at the next scheduled execution, typically granting root ownership to an interactive shell.

\noindent\textbf{Systemd service (T1543.002).} The trigger is a systemd service unit writable by a lower-privilege user while being loaded and executed by root. Exploitation rewrites the execution directive so that the next activation of the unit runs attacker-controlled code with root privileges.

\noindent\textbf{Password disclosure (T1552.001/.003).} The trigger is plaintext credentials accessible to the low-privilege user, for example in shell history, application configuration files, or environment variables. Exploitation searches common locations for candidate secrets and then authenticates as the privileged account whose credentials were recovered.

\noindent\textbf{SSH key injection (T1098.004).} The trigger is a writable \texttt{authorized\_keys} file attached to a privileged account. Exploitation appends the attacker's public key and then logs in with the matching private key.

\noindent\textbf{DB credential privesc (T1078.003).} The trigger is a database account whose password is reused by a privileged local OS account (e.g., \texttt{root} or a sudoer). Exploitation recovers the credentials from database configuration files or environment variables accessible to the low-privilege user, then authenticates as the privileged local account at the OS layer.

\noindent\textbf{Docker/container escape (T1611).} The trigger is over-broad access to the container runtime, such as docker-group membership, a mounted host filesystem, or a usable Docker socket. Exploitation launches a privileged auxiliary container that mounts the host filesystem and spawns a shell inside it, escaping the sandbox with host-root privilege.

\section{Backward Compatibility with hackingBuddyGPT}
\label{app:backward-compat}

As a secondary validation, we evaluate all models on the 13 overlapping hackingBuddyGPT scenarios to compare \textsc{PrivEscalate} results with prior work. Claude Sonnet~4.6 achieves 46.2\% SR (6/13), GPT-5.4 achieves 38.5\% (5/13), and GPT-4.1 achieves 30.8\% (4/13), consistent with Happe et al.'s reported 33 to 46\% zero-guidance SR range for GPT-4-Turbo~\cite{happe2026llms}. DeepSeek~v3.2 matches Claude Sonnet~4.6 at 46.2\% on these legacy scenarios despite lower overall SR. This provides a compatibility check between \textsc{PrivEscalate}'s automated pipeline and prior manually crafted scenarios.

\section{Expert Extension for Frontier Categories}
\label{app:expert}

The 8~expert-authored scenarios in the audited benchmark address two orthogonal gaps in the automated pipeline.

\noindent\textbf{Gap~1: Generation failure.} \Cref{tab:construction} reports zero pipeline-verified scenarios for six sub-categories (Polkit, D-Bus, PATH hijacking, LD\_PRELOAD, systemd service, DB credential reuse). To validate reachability with human authoring, we hand-craft one Dockerized scenario per category following the same interface and differential-verification protocol as the original scenarios, each implementing the canonical exploitation pattern (e.g., a permissive polkit rule, a writable systemd unit, or a sudoers entry preserving LD\_PRELOAD). All six pass the Verifier's build and differential checks (\Cref{sec:verifier}); given the small sample, per-agent success is reported as supplementary evidence.

\noindent\textbf{Gap~2: Ingestion absence.} Weak file permissions cannot be represented in GTFOBins's binary-to-exploit index because the primitive is a file-mode error rather than a binary capability, and Exploit-DB lists it only as a secondary pre-condition. We therefore hand-craft two scenarios with world-writable system credential files; these also replace the two audit-removed pipeline outputs (\Cref{sec:audit}).

\section{Construction Pipeline Details}
\label{app:pipeline-details}

This section collects supplementary construction-pipeline diagnostics deferred from \Cref{sec:construction-stats}.

\noindent\textbf{Filter and verification statistics.} GTFOBins entries flow through the pipeline without filtering: the index already provides structured binary-to-technique mappings, so the DataIngester emits one template per (binary, exploit type) pair, yielding the 771~templates reported in \Cref{tab:construction}. Exploit-DB requires four successive filters to reduce its 527~raw entries (cited in \Cref{sec:privescgen}) to 296~pipeline-input templates: (1)~deduplication removes duplicate CVE entries, leaving 499~unique; (2)~keyword-based pre-filtering drops 195~obvious kernel exploits, leaving 304; (3)~LLM-based Docker feasibility classification retains 303~Docker-feasible entries (including 9 that require elevated container privileges) and excludes 1 further entry that requires special host configuration; (4)~a final manual review removes 7~kernel exploits missed by the keyword filter, yielding the 296~templates reported in \Cref{tab:construction}.

\noindent\textbf{Verification as diagnostic tool.} The triple-verification protocol requires a successful build, root acquisition on the vulnerable image, and blocking of the same path on the fixed image. Failure counts by stage are reported in \Cref{sec:construction-stats}; requiring all three checks prevents build-only validation from masking ineffective exploits or incomplete fixes.

\section{Generative AI Usage Declaration}
\label{app:genai}

In accordance with ACM policy on authorship and the use of generative AI tools, we disclose the following uses of AI-assisted tools in the preparation of this work. For \textbf{benchmark construction}, our construction pipeline (\Cref{sec:privescgen}) uses an LLM-based agent for automated Dockerfile generation, exploit script adaptation, and verification diagnosis; these generated artifacts are the subject of study and undergo triple automated verification. For \textbf{writing assistance}, an LLM-based writing assistant was used to polish manuscript text for grammar, clarity, and phrasing, with all suggested edits reviewed and verified by the authors. For \textbf{code development}, an AI coding assistant was used to assist in code development, with all code reviewed and tested by the authors. All AI-generated content was reviewed, verified, and edited by the authors, who take full responsibility for the accuracy and integrity of the final work.

\section{System Prompts}
\label{app:prompts}

The complete system and user-turn prompts for the construction pipeline and \textsc{PrivEscAgent} are released in the artifact under \texttt{privescgen/prompts/} and \texttt{privescagent/}.

\end{document}